\documentclass[final,3p]{elsarticle}

\usepackage{booktabs}
\usepackage{pifont}
\newcommand*\rot{\rotatebox{90}}
\usepackage{rotating}
\newcommand*\OK{\ding{51}}
\usepackage{svg}

 \usepackage{array,multirow}
\usepackage{textcomp}

\usepackage[export]{adjustbox}

\usepackage{longtable}

\usepackage{caption}
\usepackage{lipsum}
\usepackage{graphicx}

\usepackage{soul}
\usepackage[colorinlistoftodos]{todonotes}

\usepackage{lscape}

\usepackage{pdflscape}

\usepackage{booktabs}
\usepackage{tablefootnote}

\usepackage[most]{tcolorbox}
\usepackage{pgfgantt}
\usepackage{tikz}
\usetikzlibrary{positioning}

\usepackage[italian]{minitoc}

\usepackage{circledsteps}

\usepackage[english]{babel}
\usepackage[utf8]{inputenc}
\usepackage[T1]{fontenc}

\usepackage{pifont}
 \usepackage{array,multirow}

\usepackage{lipsum}

\usepackage{lineno,hyperref}
\modulolinenumbers[5]

\hypersetup{
    colorlinks=true,linkcolor=black
}

\journal{Computer Science Review}

\usepackage{tocloft}

\biboptions{compress}

\begin{document}

\begin{frontmatter}

\title{Relating Edge Computing and Microservices by means of Architecture Approaches and Features, Orchestration, Choreography, and Offloading: A Systematic Literature Review}

\author{Lucas Fernando Souza de Castro\corref{cor1}%
    \fnref{fn1,fn2}}
\ead{lucas.castro@ic.unicamp.br}

\author{Sandro Rigo%
    \fnref{fn1,fn2}}
\ead{srigo@unicamp.br}

\cortext[cor1]{Corresponding author}
\fntext[fn1]{University of Campinas - Institute of Computing (IC-UNICAMP) - Av. Albert Einsten, 1251 - Cidade Universitaria, Campinas (Brazil)}
\fntext[fn2]{Preprint version - Submitted article to \textit{Computer Science Review Journal (Elsevier)}}


\begin{abstract}

\textbf{Context:} Microservices running and being powered by Edge Computing have been gaining much attention in the industry and academia. Since 2014, when Martin Fowler popularized the Microservice term, many studies have been published relating these subjects to explore how the Edge’s low-latency feature could be combined with the high throughput of the distributed paradigm from Microservices. \textbf{Objective:} Identifying how Microservices work together with Edge Computing whereas they take advantage when running on Edge.
\textbf{Method:} In order to better understand this relationship, we first identified its key concepts, which are: architecture approaches and features, microservice composition (orchestration/choreography), and offloading. Afterward, we conducted a Systematic Literature Review (SLR) as the survey method.
\textbf{Results:} We reviewed 111 selected studies and built a taxonomy of Microservices on Edge Computing demonstrating their current architecture approaches and features, composition, and offloading modes. Moreover, we identify the research gaps and trends.
\textbf{Conclusion:} This paper is a step forward to help researchers and professionals get a general overview of how Microservices and Edge have been related in the last years. It also discusses gaps and research trends. This SLR will also be a good introduction for new researchers in Edge and Microservices.
\end{abstract}


\end{frontmatter}

\renewcommand\cftaftertoctitle{\par\noindent\hrulefill\par}
\tableofcontents
\noindent\hrulefill

\section{Introduction}

Several reports point out how the data at the network's Edge has been increasing \cite{satyanarayanan_emergence_2017,edge_promise,edge_a_prime_2018,gartner_edge_report}, reaching a point where sending these data to be processed in the Cloud is no longer a viable solution. The Fog Computing concept appeared in 2012, proposing how intermediate layers between the network edge and the Cloud could be gainful in terms of lower response time, bringing a better quality of service (QoS) to the users. Fog Computing works mainly at the core network instead of far away in the Cloud data centers\cite{bonomi2012fog}.

Besides Fog Computing, another term has become popular, Edge Computing. This concept focuses on the layer next to the users (even closer than the Fog), providing them with lower response times and higher availability. However, there are many definitions of Edge Computing and what it should bring to the users in the literature. For example, the authors in~\cite{isedge_solution} assert that Edge is an abstract concept and that Fog Computing is one possible implementation of Edge Computing. On the other hand, in \cite{buyya_fog_2018}, the authors say that Edge and Fog are distinct concepts, where the Fog works in the core network, whereas the Edge is in the access network (closer to the users). This review follows the latter definition, where Fog and Edge are different concepts.

The Edge Computing scenario is composed of edge user devices (e.g., smartphones, personal computers, IoT devices) and edge server nodes (e.g., a Raspberry Pi or an Intel Galileo)~\cite{sonmez_edgecloudsim:_2017}. In the Edge, users can move, change the demand for a specific application, connect, and disconnect. Thus, applications running on the Edge must be able to adapt to these characteristics and run on edge servers. In particular, these servers usually provide fewer computational resources than those in a Cloud data center. 

Applications based on the Monolithic architecture model are known for having the presentation, business logic, and data access layer in the same entity. Nevertheless, some authors showed that whereas the application increases in features or customization, it becomes a non-trivial task to scale or maintain the system \cite{microservices_size_matters,microservices:_2017_yesterday_today_tom}. On the other hand, the Microservice architecture mode splits parts of the system (presentation, data access, logic) into small services, the so-called Microservices. Each microservice is in charge of one business capability and can communicate with its peers to synchronize, coordinate, and cooperate in providing the system’s functionalities to the users. Several projects proposed using microservices in Edge Computing due to their distributed characteristic while meeting edge servers’ resource requirements. An Edge server handles a microservice instance better than a whole monolithic application. 

Developing a Microservice-based system able to run in the Edge also brings challenges due to the dynamicity of this scenario. Microservices could be managed by a central orchestrator (or a group of) or even run in choreography mode. Also, they may offload among the Edge nodes or even to the Fog or Cloud. Moreover, microservices from different programming languages or platforms should be able to work synergistically (interoperability).  Several solutions have been designed in the literature to face these challenges. For example, simulating Edge Computing applications to verify how many edge nodes are required  \cite{egdecloudsim2020}, employing deep learning to predict the user's behavior \cite{edge_systematic_mapping_study_2021}, deploying the microservices into Docker containers and orchestrating them through Docker Swarm \cite{paper_58} or Kubernetes \cite{paper_41,paper_98}.

Different architecture features appeared over the last years to enhance and benefit from the microservices running in the Edge, in particular, to fulfill the low latency required by this environment\cite{satyanarayanan_emergence_2017}. Moreover, Edge Computing also brings many additional challenges to be solved, like security, availability, and scalability. The architecture features propose to reduce (or even solve) the challenges that Edge Computing brings. For instance, some solutions use the blockchain to track the microservices’ transactions to improve security and node reliability. Further, the API Gateway is another example of an architectural feature employed in many solutions. As far as the approaches, the architectures may use a single~\cite{paper_24,paper_26}, two~\cite{paper_10,paper_78}, or even all three \cite{paper_58,paper_103} tiers (Edge, Fog, Cloud) in the architecture to host and run the microservices to achieve the expected QoS~\cite{paper_58,paper_103}. Other authors proposed extra tiers beyond the traditional 3-tier approach. These tiers could be below (Mist, Dew\cite{paper_27}) or above the Edge, and they serve the Edge/IoT devices and provide services to the Edge (e.g., an AI service~\cite{paper_103}). Besides the tier aspect, the architecture features also focus on design patterns, tools, frameworks, and programming languages. 

Let us dive deeper into microservice architectures. Two main aspects are essential to work in Edge Computing: microservice composition and computation offloading. Microservice composition is how the microservices are arranged and managed to answer incoming requests. There are mainly two ways of composing microservices: orchestration and choreography \cite{microservices:_2017_yesterday_today_tom,microservicespatternsBook}. Orchestration is a centralized way of managing microservices by employing a single orchestrator~\cite{alam_orchestration_2018,microservices_size_matters}. Moreover, some container-based architectures use their container orchestrator (e.g., Docker Swarm or Kubernetes) to orchestrate the microservices \cite{paper_58}. However, several authors argue that specific microservice orchestrators are necessary to handle the dynamicity of this environment. That would generate more robust and flexible microservice architectures for Edge Computing \cite{paper_38,paper_95,paper_102}. 

Computation offloading is a crucial feature in Edge Computing \cite{edge_promise,isedge_solution}, where heavy tasks can offload from the Edge to a higher tier (e.g., Cloud) to take advantage of a bigger computational power. Moreover, some tasks may run in the Edge while others run in the Cloud, and the architecture must handle them concomitantly. Whenever the task in the Cloud gets done, it offloads back to the Edge to deliver its outcome. There is a key question to be answered: what can be offloaded? The authors in \cite{paper_46} describe how to offload a code, a VM, or a thread. Thus, a new question arises: could we offload microservices? A microservice could be offloaded to a different node in the same tier (horizontal) or even on another tier (vertical). The offloading is mainly motivated by performance issues but also could be impelled by geographic constraints, network failures, architecture policy, or topology switches \cite{paper_87,paper_22,paper_38}. Finally, there are three main aspects to consider to build a microservice-based architecture for Edge Computing: the architecture approach and features, the microservice composition, and the computation offloading. The architectural 99approach requires considering other important aspects by itself. We give further details in the following sections. 

In order to track these aspects relating the Microservices and Edge Computing, we have conducted a Systematic Literature Review (SLR) based on guidelines from Kitchenham and Charters \cite{kitchenham2007slr,kitchenham2015evidence}. Our SLR analyzed a total of 1818 papers from well-known academic databases according to a designed research string. After three selection steps, we collected 111 papers. Through 13 research questions, a deeper analysis verified the solutions employed by relating Edge Computing and Microservices in terms of Architecture Approach, Microservice Composition, and Offloading. 

The outcomes from the SLR are a step forward in developing a detailed view of how we compose, offload, and architect the microservices when they work in Edge Computing. Every research question brings along a discussion addressing how the selected studies deal with the subjects. Moreover, we identify the trends pointed out by the selected studies as well as the research gaps. In addition, we built a broad taxonomy to summarize the trends and illustrate the research directions taken by the selected studies. 

The rest of the paper is organized as follows. Section \ref{sec:theoretical} describes the theoretical background of Edge Computing and Microservices and their related concepts. Section \ref{sec:slr_met} introduces the SLR steps and shows how we conducted them. Section \ref{sec:slr_results} presents the SLR results. Finally, Section \ref{sec:conclusion} draws our conclusions.

\section{Theoretical Background}
\label{sec:theoretical}

This section briefly discusses the key concepts explored in this literature review. In particular, Edge Computing and the main aspects of Microservice Architectures.

\subsection{Edge Computing}

The search for solutions to the bottleneck between the user at the network edge and the server mainly running in the cloud is not novel. However, the Edge Computing terminology is somewhat recent.

This network bottleneck started to be noticed in the 2000s when IoT devices and mobile computing became popular \cite{satyanarayanan_emergence_2017}. Thus, how could this enormous amount of IoT devices be connected in a very low-latency network and with good speed? In \cite{satyanarayanan2009case}, the authors showed how the increasing WAN delay and jitter affect the overall user experience when running interactive tasks (e.g., virtual reality, video streaming). Their proposed solution was the creation of cloudlets, a "datacenter in a small box near to the users \cite{satyanarayanan2009case,satyanarayanan_brief_2015}". Moreover, a former Cisco report pointed out that circa 2022 there would be around 50 billion IoT devices worldwide \cite{evans_cisco}. The question that Cisco raises is: "How to guarantee low WAN latency and jitter to this huge amount of IoT devices worldwide while providing computational resources to fulfill the mobile user requirements simultaneously?". 

Following up on the cloudlets in \cite{satyanarayanan2009case,satyanarayanan_brief_2015}, Bonomi et al. ~\cite{bonomi2012fog} proposed the concept of Fog Computing. The idea behind Fog Computing is to provide hierarchical layers between the cloud and the edge network. These layers are called the Fog. In the Fog, multiple servers provide computational resources to the users. 

Fog Computing designs, improvements, and extensions explore these hierarchical layers regarding security, availability, multiple access, services, and so on \cite{buyya_fog_2018}. Some of these extensions include Mist, Osmotic, and Edge Computing.

\begin{figure*}[h]
    \centering
    \fbox{\includegraphics[width=.97\textwidth]{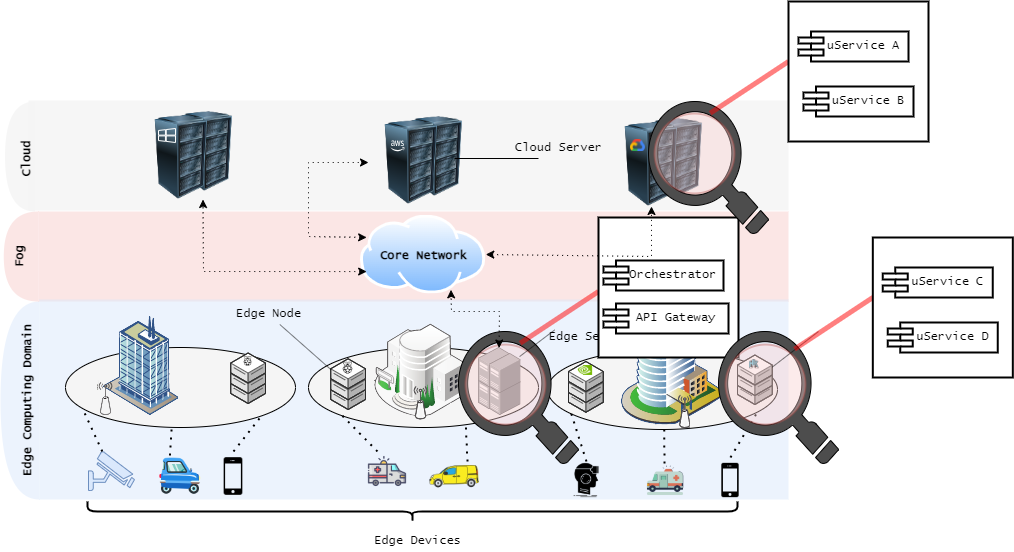}}
    \caption{Computational domain of Cloud, Fog, and Edge Computing employing Microservices}
   	\label{fig:buyya_fog}
\end{figure*}


The Edge Computing proposal is similar to Fog Computing: "Provide computational resources closer to users with a lower latency than the Cloud Computing model" \cite{satyanarayanan_emergence_2017,edge_promise,edge_visionChallenges_Shi}. The main difference between Edge and Fog is how they are designed and deployed. Figure~\ref{fig:buyya_fog} illustrates the taxonomy of Cloud, Fog, and Edge Computing computational models and their variations. 

There are several definitions in the literature for Edge Computing, some of which are controversial. For example, the authors in \cite{comparison} define Fog Computing as an implementation of Edge Computing. On the other hand, the authors in \cite{buyya_fog_2018, edge_a_prime_2018} define them as distinct and independent concepts. Thus, for the work presented here, we will adopt that Edge Computing consists of a model whose purpose is to approximate the resource processing and provisioning to the edge of the network, that is, the closest to the user. As such, Edge Computing is one hop away from the user.

\subsection{Microservices}

The architecture model of monolithic systems consists of developing applications where the presentation, logic, and persistence layers are in a single block. This model is feasible for small solutions since there is no great complexity in the layers and their interactions. However, as systems grow and acquire new functions, the complexity of development and maintenance becomes increasingly higher. The component-oriented development methodology and service-oriented architectures \textit{(SOA - Service Oriented Architecture)} \cite{microservices:_2017_yesterday_today_tom} are examples of solutions existing in the literature. 

Unlike service-oriented architectures, microservice architectures have a finer granularity of their modules (microservices) and less complexity in exchanging messages between them. Their modeling philosophy is: 
"Do just one thing and do it well." As reported in \cite{comparative_monolith_microservices, samarpit_tuli_microservices_2019}, SOA services grow significantly in some cases, reaching the status of systems or sub-systems. 

In summary, microservices could be defined as a \textit{"cohesive, independent process interacting via messages"} \cite{microservices:_2017_yesterday_today_tom} and is an \textit{"architectural style stemming from Service-Oriented Architectures, ... it is a small building block that communicates only through message passing"} \cite{microservices_size_matters}. The microservice architectural style \textit{"is a distributed application where all its modules are microservices"} \cite{microservices:_2017_yesterday_today_tom}. 

The main microservice characteristics that differ from monolithic and service-oriented systems are in the following three principles \cite{microservices:_2017_yesterday_today_tom,comparative_monolith_microservices}:

\begin{enumerate}
    \item \textbf{Bounded Context:} Services implement the business capabilities. A single service encapsulates all the strongly related functionalities; 
    \item \textbf{Size:} It is smaller than the services from the Service Oriented Architecture. This feature is essential to the microservice architecture. If a microservice is large, it must be split into two different microservices to keep the bounded context feature; 
    \item \textbf{Independence:}  Each microservice should be independent in terms of having its own database, logic, and business layer \cite{microservice_designpatterns}. However, the microservice can communicate with its peers by sharing a common interface. 
\end{enumerate}

Microservice architectures are simpler to develop and maintain and present advantages in terms of performance and scalability. For example, some experiments conducted by researchers at IBM found that systems developed using microservices have a higher yield when demanded by many requests \cite{ibm}. 

\subsubsection*{Architecting Microservices}

Architecting systems through microservices brings a higher level of complexity than a regular distributed application. Designers must plan how the microservices will interact to bring scalability, maintainability, and performance to the architecture. Below, we describe some architectural features that microservices could provide to compose a robust architecture. 
 

\begin{itemize}
   \item \textbf{Microservice Composition Modes:} The user views a system built through microservices as a unique and single system. However, this system is a set of microservices working together to bring all its functionalities to the users. Thus, defining the architectural approach to compose the microservices is critical to their design. There are two major composition approaches: Orchestration and Choreography. 

    \begin{itemize}
        \item \textbf{Orchestration:} This mode requires an orchestrator, a conductor of the microservices. Some architectures suggest that the orchestrator could also be a microservice \cite{microservices_size_matters,paper_8}. In this mode, the architecture works in a centralized way. The orchestrator coordinates the functionalities of different microservices and combines the outcomes. Figure \ref{fig:microserviceCompositions} shows an architecture in the orchestration composition mode (left side). 
        
         \begin{minipage}{\linewidth}
            \centering
            \fbox{\includegraphics[width=0.97\textwidth]{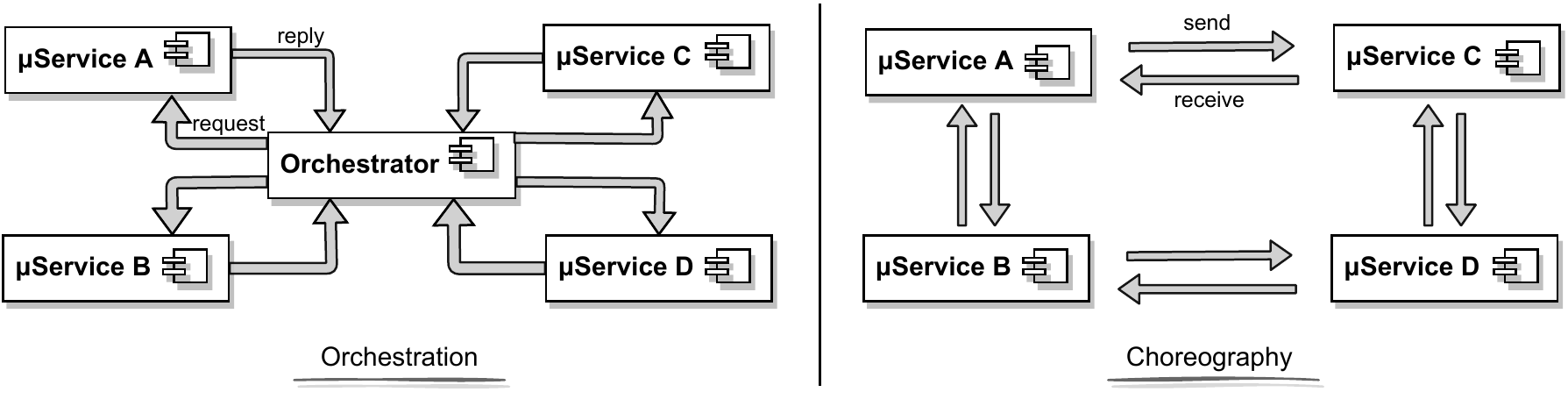}}
            \captionof{figure}{Microservice Compositions - choreography and orchestration - Adapted from \cite{disambiguation_2017_micro}}
           	\label{fig:microserviceCompositions}
        \end{minipage}
        
        \item \textbf{Choreography:} The choreographic mode does not use an orchestrator. The coordination of the microservices is distributed among all microservices. They coordinate by passing messages and establishing rules and agreements to provide the system functionalities Figure \ref{fig:microserviceCompositions} demonstrates an architecture in the choreography composition mode (right side) 

In his book \textit{Building Microservices}, Saw Newman makes a valuable comparison between Orchestration and Choreography:        

\begin{quote}

\textit{``With orchestration, we rely on a central brain to guide and drive the
process, much like the conductor in an orchestra. With choreography, we inform
each part of the system of its job, and let it work out the details, like dancers all finding their way and reacting to others around them in a ballet.''} \cite{buildingmicroservices_samnewman_book} 
\end{quote} 
    
    \end{itemize}
    \item \textbf{Workflows:} In the ordinary meaning, the workflow concept refers to tasks, procedures, actors, input, and output to coordinate a business process describing it step-by-step. The term appeared in the 20s in the manufacturing industry, where workflows described the process in the assembly line \cite{workflowbook}. Since the workflow comprises steps, it may be a diagram, which aims to demonstrate three main components: (1) input description, (2) transformation rules, and (3) output description \cite{microserviceJolie}. Different tools can describe this process, for instance, the Business Process Model and Notation (BPMN) \cite{chinosi2012bpmn}.

    The workflow describes how a business process works along with the system and its components. For example, in \cite{paper_16}, the authors define a workflow to specify how microservices must conduct an image extraction. Furthermore, to manage the several workflows that a microservice architecture can handle, some architectures use a Workflow Engine, which is a mechanism that drives task delegation, task chaining, flow timeline, and deadlines \cite{workflowsite_microservice_cambda}. Hence, the Workflow Engine is crucial in coordinating the Microservices in these architectures. In order to illustrate how the Workflow Engine works, we took as an example the workflow presented in \cite{paper_16} and expanded it (Figure \ref{fig:workflowExample}) to show the Workflow Engine tasking delegation. 
    
        \begin{figure*}[h]
            \centering
            \fbox{\includegraphics[width=0.75\textwidth]{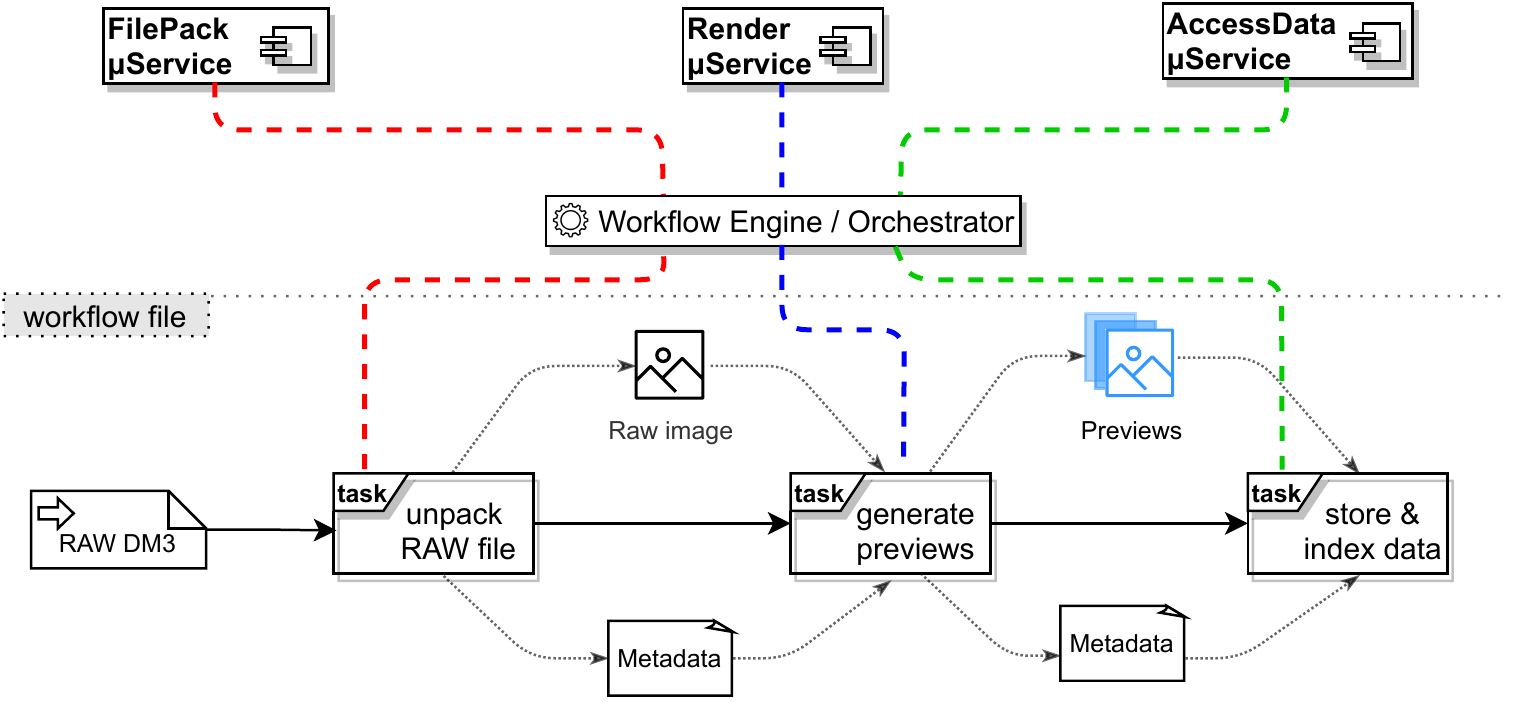}}
            \caption{Workflow Example - Adapted and expanded from \cite{paper_16}}
           	\label{fig:workflowExample}
        \end{figure*}

    In Orchestrated Microservice Architectures, the orchestrator can perform the tasks that the Workflow Engine provides \cite{workflowsite_microservice_cambda}. Moreover, the Workflow Engine could also be employed in Choreography Microservice Architectures. However, it does not work as a centralized entity.
    
   Finally, specifying a workflow in a microservice architecture is important to guarantee how microservices perform tasks. The workflow helps the orchestrator to manage how tasks are conducted and map them if any task or transaction is not acting as expected and designed in the workflow. In choreography architectures, the workflow helps each microservice to ensure its tasks and operations cooperate with its peers.

    \item \textbf{Offloading:}

    The microservice offloading\footnote{Some authors employ this concept by using different therms, such as: i) microservice dynamic placement \cite{paper_94} or replacement \cite{paper_104}; ii) dynamic deployment and re-deployment \cite{paper_65,paper_80}; iii) microservice migration \cite{paper_84,paper_61}} works upon the computation offloading concept. Microservice offloading may be demanded due to several reasons: (1) demand change, (2) host node (or container) resources running low, and (3) security and privacy. Furthermore, depending on the application, the microservice does not offload as a whole but only as a task, a thread, or a workload \cite{paper_6,paper_45}. Figure \ref{fig:offloadExample} shows an example where an orchestrator offloads a Microservice ($\mu$Service A running in Node A) to another host (Node B). The offloading process is requested because Node A is running low on computational resources. 
    
       \begin{figure}[h]
            \centering
            \fbox{\includegraphics[width=0.6\textwidth]{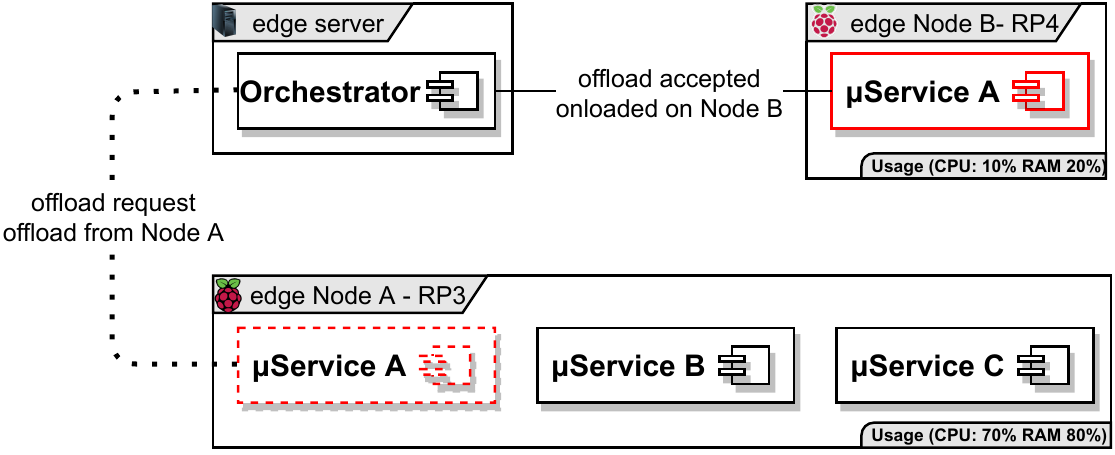}}
            \caption{Offloading Example}
           	\label{fig:offloadExample}
        \end{figure}
        
    There are two primary modes of offloading in a microservice architecture: horizontal and vertical. The horizontal mode provides the offloading in the same tier (layer). For example, in Edge Computing, horizontal offloading is the offloading among the Edge Nodes. On the other hand, vertical offloading is when a microservice offloads among different tiers (Edge to Fog to Cloud, Fog to Cloud, Edge to Cloud) \cite{paper_1,paper_31}.
    
\end{itemize}

\section{The Systematic Review - Methodology, Steps and Setup}
\label{sec:slr_met}

\begin{figure}[h!]
    \centering
    \fbox{\includegraphics[width=0.975\textwidth]{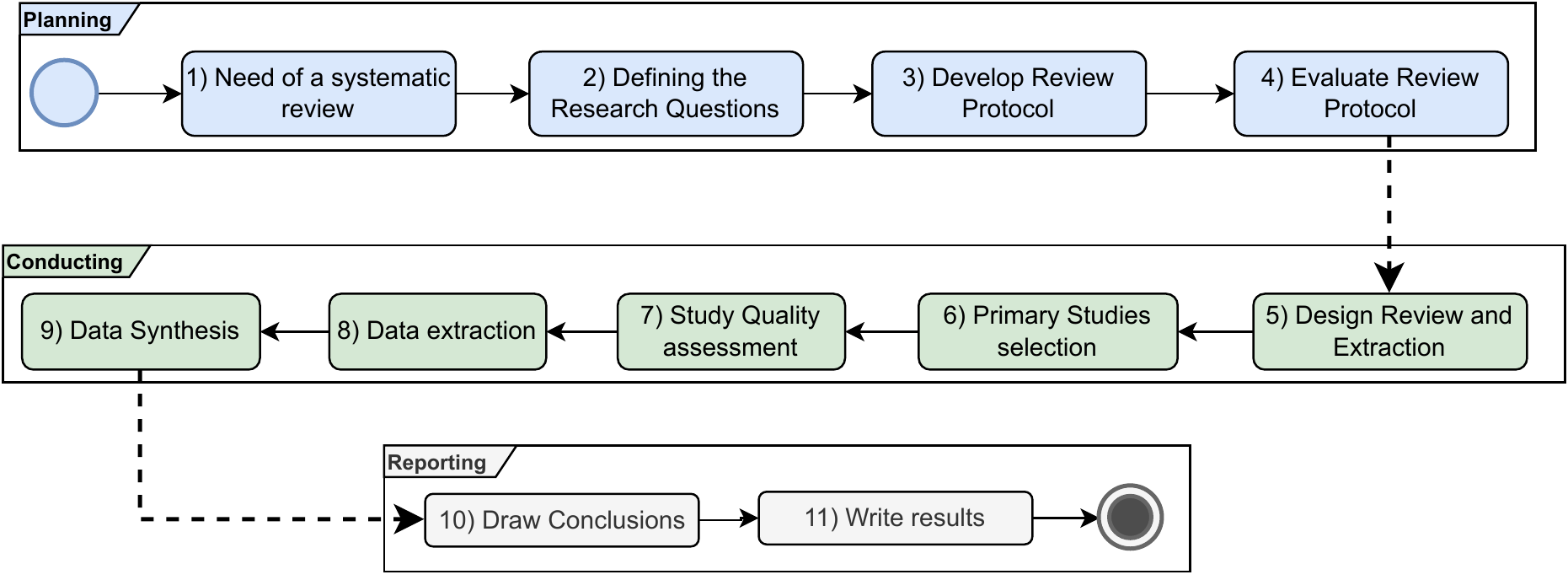}}
    \caption{SLR step by step}
   	\label{fig:slrsteps}
\end{figure}

We adopted the Systematic Literature Review  (SLR) method based on Kitchenham and Charters 2007 guidelines \cite{kitchenham2007slr}. The authors propose to split the revision into three major phases: Planning, Conducting, and Reporting. Figure \ref{fig:slrsteps} illustrates these phases and their sub-phases.

The \textit{Need for a systematic review} (as seen in Figure \ref{fig:slrsteps} - Step 1) was motivated mainly due to a vast amount of new research and projects relating Edge Computing with Microservices. Although the concept of Edge Computing was coined in the 1990s, it was after 2015 that it gained popularity \cite{edge_systematic_mapping_study_2021}. Lewis and Fowler \cite{lewis2012microservices,microservicesiotapproach} defined the Microservice architecture in 2014. Since this concept appeared simultaneously with the increase in Edge Computing popularity, many works have proposed using microservices in Edge Computing so far. 

As far as our SLR goes, its primary goal is: tracking and analyzing the works that relate Edge Computing and Microservices in terms of Architecture Approach, Orchestration, Choreography, and Offloading.

In order to check if an SLR with the same goal as ours already existed, we performed searches on Google Scholar, Web of Science, Scopus, and other scientific databases\footnote{ACM, SpringerLink, IEEEXplore, MDPI, Elsevier ScienceDirect, and ResearchGate}. As a result, previous SLRs still needed to fulfill our goals.

\subsection{Planning - Research Questions}

\begin{figure*}[h!]
    \centering
    \fbox{\includegraphics[width=.975\textwidth]{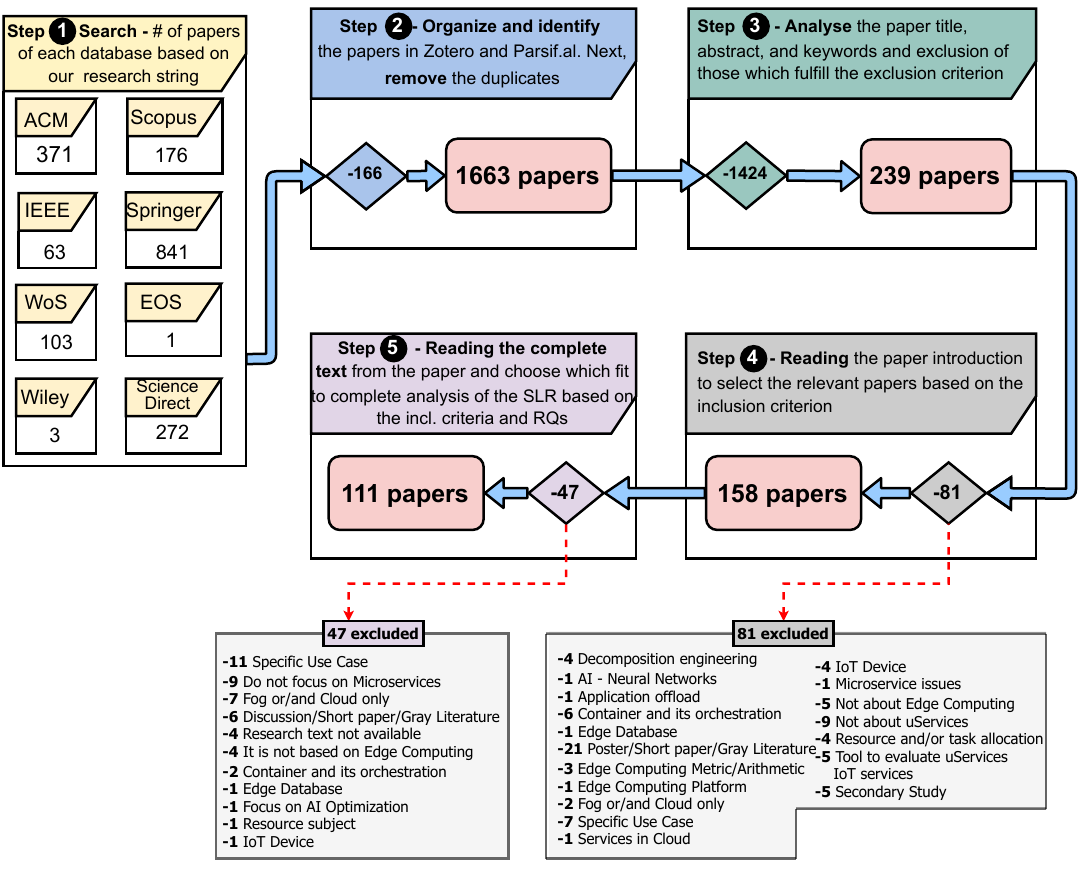}}
    \caption{SRL - Papers Overview}
   	\label{fig:papersoverview}
\end{figure*}

Besides the terms Architecture Approaches, Orchestration, Choreography, and Offloading, we designed specific research questions to identify which works relate to and fill the gaps between Edge Computing and Microservices. Thus, our target is to answer every single question in Table \ref{tab:researchQuestions}.

\begin{table*}[ht]
\centering
\begin{tabular}{@{}ll@{}}
\toprule
\textbf{Research Questions}                                                                                                                                                                      & \textbf{Description and motivation}                                                                                                                                                                      \\ \midrule
\begin{tabular}[c]{@{}l@{}}
RQ 1. What are the existent \\
microservice architectures' \\
approaches and features to \\
Edge Computing?\end{tabular}                                                           & \begin{tabular}[c]{@{}l@{}}
Identify and analyze the existent microservice approaches and \\
features in the literature that uses Edge Computing as its\\
application scenario. \end{tabular}                                                                                                                                                                                     \\ \midrule
\begin{tabular}[c]{@{}l@{}}
RQ 1.1 Based on the microservice \\
architectural approaches and \\
features, are there advantages of \\
using microservices in the Edge?\end{tabular} & \begin{tabular}[c]{@{}l@{}}
Identify and list the microservices' advantages and \\
disadvantages in the Edge Computing scenario. \end{tabular}
\\ \midrule
\begin{tabular}[c]{@{}l@{}}
RQ 1.2. From the architectural \\
perspective, which microservice \\
characteristics take advantage of \\
Edge Computing usage? \end{tabular}           & \begin{tabular}[c]{@{}l@{}}

Analyze specific microservice features and their details that \\
take advantage of or even gain performance when running in Edge \\
Computing. Some microservice features, like reactivity, are key \\
to their success. Thus, how do microservices behave when reacting \\
in the Edge? 
\end{tabular}                      \\ \midrule
\begin{tabular}[c]{@{}l@{}}
RQ 2. What are the fundamental \\
microservice characteristics to \\
properly work in Edge Computing?\end{tabular}                                & \begin{tabular}[c]{@{}l@{}}

Verify the crucial microservice features to work in the Edge \\
Computing scenario with a QoS guaranteed. For example, in \\
Edge Computing, low latencies are a prerequisite. Hence, \\
what microservice feature fulfills this? 

\end{tabular}                                                              \\ \midrule
\begin{tabular}[c]{@{}l@{}}
RQ 3. What are the techniques\\
employed in the microservice\\
composition process? \\
In which tiers do they work?\end{tabular}                                        & \begin{tabular}[c]{@{}l@{}}
The microservice composition could work in two ways: orchestrated \\ 
or choreography. Thus, this question aims to identify the \\
existent composition techniques and tools that microservices work \\
upon. Moreover, in which tiers or layers (edge, fog, and cloud) do\\
these techniques manage?\end{tabular} \\ \midrule
\begin{tabular}[c]{@{}l@{}}
RQ 4. How does the microservice \\
offloading process work? 
\end{tabular}                                                                         & \begin{tabular}[c]{@{}l@{}}
Identify and analyze the current techniques and tools that \\
support microservice offloading among the edge nodes \\
(horizontally) and tiers (vertically).\end{tabular}                                                                                                                                                                            \\ \midrule
\begin{tabular}[c]{@{}l@{}}
RQ 5. What challenges/problems\\
are identified in research \\
literature relating to microservices\\
and Edge Computing?\end{tabular}                                  & \begin{tabular}[c]{@{}l@{}} 

What are the challenges that Microservices and Edge Computing \\
impose on the researchers? And, what are the needed works? 

\end{tabular}                                                                                                                    
\\ \bottomrule
\end{tabular}
\caption{Research Questions}
\label{tab:researchQuestions}

\end{table*}

\subsection{Planning - Review Protocol}
\label{subsubsec:reviewprotocol}

The review protocol has three main phases: i) Database selection; ii) Search String Design; iii) Specify the inclusion and exclusion criteria. 

Firstly, we chose multi-disciplinary scientific databases that are relevant to computer science. The databases were: ACM Digital Library, IEEEXplore, Web of Science (WOS), Scopus, SpringerLink, and ScienceDirect. Besides their relevance, these databases also provide advanced search support. \footnote{We used the database aggregators (Web of Sicence and Scopus). However, we did not reference the aggregators as the source of the selected studies. Instead, we always used the original databases as the source. For instance, some papers were listed by Web of Science, but were published on an IEEE magazine. Thus, the source was recorded as IEEE.}

Secondly, we designed and built the search string. This string was employed to search the paper’s title, abstract, and keywords. It was constructed iteratively through several attempts to retrieve relevant documents in the databases.

\begin{figure}[h!]
    \centering
    \begin{tabular}{|c|}
        \midrule
        \textit{microservice* AND (edge computing OR "edge clouds" OR osmotic computing) AND}\\ \textit{(orchestrat* OR architectur* OR choreography OR offload*)}\\
        \midrule
    \end{tabular}
    \label{fig:researchString}
    \caption{Research String}
\end{figure}

The subjects’ precedence and relevance to the SLR determine the keyword order in the search string. For instance, we are more focused on papers that explicitly use microservices. Moreover, we are looking for papers that focus the microservices in Edge Computing, Osmotic Computing, and even in Edge Clouds. We include Osmotic Computing and Edge Clouds for two main reasons: i) Osmotic Computing combines Edge and Cloud Computing. Several authors focus more on Edge than Cloud in this paradigm. ii) Edge Clouds is a definition some authors use to specify the Edge Computing scenario. Finally, the \textit{architectur*, orchestrat*, choreography, and offload*} keywords were included. We did not use the microservice composition keyword because it may appear in several forms: orchestration, choreography, or even as microservice coordination. Therefore, we exhaustively tried to find a broad and easy search string.

Finally, we must define the inclusion and exclusion criteria to accept or reject a paper. We are only interested in primary studies published between 2016 and 2022 (up to May)\footnote{This SLR step was performed in May 2022} and those that brought relevant contributions. Table  \ref{tab:inclusionCriteria}  shows the inclusion-exclusion criteria employed in the SLR.

\begin{table}
\centering
\begin{tabular}{@{}lll@{}}
\toprule
\textbf{\#} &
  \textbf{Inclusion Criteria} &
  \textbf{Exclusion Criteria} \\ \midrule
0 &
  \begin{tabular}[c]{@{}l@{}}Primary studies that were published\\ between 2016 and 2022*\end{tabular} &
  \begin{tabular}[c]{@{}l@{}}Gray literature, Secondary Studies\\ Older than 2016\end{tabular} \\ \midrule
1 &
  \begin{tabular}[c]{@{}l@{}}Microservice Architectures on Edge Computing\end{tabular} &
  \begin{tabular}[c]{@{}l@{}}Research text not available\end{tabular} \\ \midrule
2 &
  \begin{tabular}[c]{@{}l@{}}Orchestrating Microservices on Edge Computing\end{tabular} &
  Redundant paper \\ \midrule
3 &
  \begin{tabular}[c]{@{}l@{}}Offloading Microservices on Edge Computing\end{tabular} &
  \begin{tabular}[c]{@{}l@{}}Cloud/Fog only related\end{tabular} \\ \midrule
4 &\begin{tabular}[c]{@{}l@{}}
  Orchestration on Edge Computing that\\ relates Microservices\end{tabular} &
  \begin{tabular}[c]{@{}l@{}}Machine Learning only focused\end{tabular} \\ \midrule
5 &
  Related to Edge Computing &
  NFV related \\ \midrule
6 &
  Related to Microservices &
  \begin{tabular}[c]{@{}l@{}}Paper focusing  only on container orchestration\end{tabular} \\ \midrule
7 &
  \begin{tabular}[c]{@{}l@{}}Studies that address and explore the relation \\between Microservices and Edge Computing\end{tabular} &
  \begin{tabular}[c]{@{}l@{}}Microservice placement\end{tabular} \\ \midrule
8 &
   &
  \begin{tabular}[c]{@{}l@{}}Focus on Service (SOA) or web-service\end{tabular} \\ \midrule
9 &
   &
  \begin{tabular}[c]{@{}l@{}}Studies clearly irrelevant to the research, \\taking into account the research questions\end{tabular} \\ \bottomrule
\end{tabular}
\caption{Inclusion and Exclusion Study Criteria}
\label{tab:inclusionExclusionCriteria}
\end{table}

\subsection{Conducting - Selecting Primary Studies}

We conducted the review and extraction based on several SLRs already published in the literature \cite{slr_fog,slr_mas_patterns,slr_vilela2017integration}. This phase (Step 5 - \ref{fig:slrsteps}) consists in developing the Quality assessment and the Data Extraction form and applying them to the selected papers. We describe Quality assessment in Section \ref{subsubsec:qualityassessment} and the Data Extraction form in Section \ref{subsubsec:dataextraction}.

Regarding selecting the papers, our process consists of five iterative steps, from the database searches to the final paper selection. Figure \ref{fig:papersoverview} illustrates these steps.

\begin{itemize}
    \item Step \Circled{\textbf{1}} - Search in databases:  ACM returned 371 titles, Scopus 176, IEEEXplore (IEEE) 63, Springer- Link (Springer) 841, Web of Science (WoS) 103, EOS 1, Wiley 3, and Science Direct 272. Therefore, we got 1818 papers in total; 
    
    \item Step \Circled{\textbf{2}} - Organize and identify: The 1818 papers were organized using Parsifal\cite{parsifalwebsite} and Zotero\footnote{Parsifal is employed to support the SLR phases and Zotero was used to store the paper files}\cite{puckett2011zotero}. In this step, we removed 166 duplicated studies, keeping 1663 papers; 
    
    \item Step \Circled{\textbf{3}} - Analysis:  We analyzed the papers' titles, keywords, and abstracts to identify and exclude studies based on the nine exclusion criteria (see Table \ref{tab:inclusionExclusionCriteria}). We removed 1424 entries. So, 239 papers proceeded to the next step;
    
    \item Step \Circled{\textbf{4}} - Introduction reading: We analyzed the introduction of the 239 papers to identify studies that meet the inclusion criteria. We excluded 81 papers, and 158 proceeded. We detail the exclusion reasons in the right frame of Figure \ref{fig:papersoverview}. 
    
    \item Step \Circled{\textbf{5}} - Complete Reading: In the last step, 158 papers were completely read to identify those which fulfill the inclusion criteria and also answer the research questions. After this step, 38 papers were removed. Finally, 111 papers were selected as relevant papers to proceed to the following stages of the SLR process, Study Quality assessment and Data Extraction. 
\end{itemize}


\subsection{Conducting - Quality Assessment}
\label{subsubsec:qualityassessment}

Quality may be a broad term. We understand quality in this SLR as how a given paper contributes towards a better understanding of the relationship between Edge Computing and Microservices. 

The quality assessment is an essential task in the SLR~\cite{kitchenham2007slr,kitchenham2015evidence}, which aims to: (i) provide more details on the inclusion-exclusion criteria; (ii) explain which quality differences each study presents in order to figure out why their results differ; (iii) provide a metric to rank the studies according to their quality; (iv) assess the research impact and, based on that, identify research gaps. 

\begin{table*}[h]
\centering
    \begin{tabular}{@{} cl*{5}c @{}}
        & & \multicolumn{5}{c}{} \\[2ex]
        & & \rot{Evaluation Research} & \rot{Validation Research} & \rot{Solution Proposal} & \rot{Experience Report} & \rot{Opinion Paper} \\
        \cmidrule[1pt]{2-7}
        & \begin{tabular}[c]{@{}l@{}} 1) Has this study increased the know-how and knowledge about \\the relationship between edge computing and microservices? \end{tabular} & \OK & \OK  &  \OK & \OK  & \OK \\ \cmidrule{2-7}
        & 2) Is there a clear statement of the goals of the research? & \OK & \OK &  & \OK &  \\ \cmidrule{2-7}
        & 3) Is the proposed technique clearly described? & \OK &   & \OK  &   & \\ \cmidrule{2-7}
        & 
        \begin{tabular}[c]{@{}l@{}}
        4) Is there an adequate description of the context (industry, laboratory \\
        setting, products used and so on) in which the research was carried out? \end{tabular} & \OK &  & \OK & &  \\ \cmidrule{2-7}
        & 5) Was the data analysis sufficiently rigorous? & \OK &  & \OK  &   &    \\ \cmidrule{2-7}
        & 6) Is there a discussion about the results? & \OK &  \OK & \OK  &   &    \\ \cmidrule{2-7}
        & 7) Are the limitations explicitly discussed? & \OK & \OK  &  \OK  &   &    \\ \cmidrule{2-7}
        \rot{\rlap{~\textbf{Questions}}}
        & 8) Are the lessons learned interesting? &  &   &   & \OK  &  \OK  \\ \cmidrule{2-7}
        & 9) Is there sufficient discussion of related work? & \OK &  \OK & \OK  &   &  \OK  \\ \cmidrule{2-7}
        & 10) Is it likely to provoke discussion? & \OK &  \OK & \OK  &   &  \OK  \\ \cmidrule{2-7}
        & 11) How well has diversity of perspective and context been explored? &  &   &   &   &  \OK  \\ \cmidrule{2-7}
        & \begin{tabular}[c]{@{}l@{}}12) How clear are the assumptions/theoretical perspectives/values \\that have shaped the form and opinions described? \end{tabular} &  &   &   &   &  \OK  \\ \cmidrule{2-7}
        & \begin{tabular}[c]{@{}l@{}}13) Do the authors explain the edge computing concept clearly? \end{tabular} & \OK &  \OK &  \OK & \OK  &  \OK  \\ 
        
        \cmidrule[1pt]{2-7}
        
    \end{tabular}
    \caption{Quality Assessment Criteria}
    \label{tab:qualityAssessment}
\end{table*}

In order to evaluate the 111 selected studies in terms of quality, we built 13 quality assessment criteria. Each criterion is judged based on the study category. We adopted the categories proposed by Wieringa et al. ~\cite{wieringa2006requirementsSLR}: Evaluation Research, Validation Research, Solution Proposal, Experience Report, and Opinion (Philosophical) Paper~\footnote{Although some SLR consider they in different categories, we considered the same category the Opinion Paper and Philosophical Paper due to their similarities.}. Table \ref{tab:qualityAssessment} introduces the quality assessment criteria. Each criterion is evaluated based on three possible scores: Yes (1), Partially (0.5), and No (0). Each study was assessed and ranked based on the sum of these grades. After this evaluation, the average quality assessment rate was 81.77\%.

\subsection{Conducting - Data Extraction}
\label{subsubsec:dataextraction}

The extraction form records and labels the information from the selected studies, retrieved in the Primary Studies selection (Step 6 - See Figure \ref{fig:slrsteps}). Every field in the extraction form has a related and relevant Research Question (RQ). Additionally, the extraction form is employed to avoid research biases, assuring that every study is evaluated (quality assessment) and scanned (data extraction) using the same criteria. Table ~\ref{tab:extractionForm} shows the extraction form.

\begin{table*}[h]
\centering
\begin{tabular}{@{}llll@{}}
\toprule
\textbf{\#} & \textbf{Study Data}                                                                       & \textbf{Description}                                                                                                                                                                                                                                                                                   & \textbf{Relevant RQ} \\ \midrule
1           & Study ID                                                                                  & Unique identifier for the study                                                                                                                                                                                                                                                                        & Study Overview       \\  \midrule
2           & \begin{tabular}[c]{@{}l@{}}Authors, Year, \\Country,  and Title\end{tabular}              & Basic article information                                                                                                                                                                                                                                                                              & Study Overview       \\  \midrule
3           &\begin{tabular}[c]{@{}l@{}} Source and\\ Article Type  \end{tabular}                                                                & \begin{tabular}[c]{@{}l@{}}
Which database does the article come from (ACM,\\
Springer, IEEE, Science Direct ,Web of Science, \\
Scopus). Also, what is the type of the article \\
(Evaluation Research, Solution Proposal, Validation \\
Research, Experience Report, or Philosophical Paper)\end{tabular} & Study Overview       \\  \midrule
4           &\begin{tabular}[c]{@{}l@{}} Evaluation \\ Method \end{tabular}                                                                         & \begin{tabular}[c]{@{}l@{}}
Which evaluation method is used:\\
Controlled experiment, case study, survey, ethnography, \\
action research, illustrative scenario, or not applicable.\end{tabular}                                                                         & Study Overview       \\  \midrule
5           & \begin{tabular}[c]{@{}l@{}}Arch. Approaches \\ and Features     \end{tabular}                                                              & \begin{tabular}[c]{@{}l@{}}
What are the approaches and features that the \\ microservice architectures demonstrated and proposed?
\end{tabular}                                                                                                                                                      & \multicolumn{1}{c}{RQ 1}                 \\  \midrule
6           & \begin{tabular}[c]{@{}l@{}}Benefits and \\ Drawbacks      \end{tabular}                                                              & \begin{tabular}[c]{@{}l@{}}What benefits and drawbacks of $\mu$Services in the EC were \\ identified in this article.\end{tabular}                                                                                                                                                                    & \multicolumn{1}{c}{RQ 1.1}               \\  \midrule
7           & \begin{tabular}[c]{@{}l@{}}
Microservice \\features taking \\ the EC Advantages\end{tabular} & \begin{tabular}[c]{@{}l@{}}What $\mu$Service features were identified by the authors that\\ take advantage of Edge Computing.\end{tabular}                                                                                                                                                             & \multicolumn{1}{c}{RQ 1.2}               \\  \midrule
8           & \begin{tabular}[c]{@{}l@{}}Essentials $\mu$Service \\ features to EC\end{tabular}         & \begin{tabular}[c]{@{}l@{}}The fundamental $\mu$Service features that were identified or\\ suggested by the authors.\end{tabular}                                                                                                                                                                      & \multicolumn{1}{c}{RQ 2}                 \\  \midrule
9           & \begin{tabular}[c]{@{}l@{}}Orchestration \\ techniques  and \\ tools/methods\end{tabular}     & \begin{tabular}[c]{@{}l@{}}
What are the orchestration tools and techniques \\
used to orchestrate the Microservices along the Edge \\
Computing Scenario?\end{tabular}                                                                                                                       & \multicolumn{1}{c}{RQ 3}                 \\  \midrule
10          & \begin{tabular}[c]{@{}l@{}}Offload techniques and \\ tools/methods\end{tabular}           & \begin{tabular}[c]{@{}l@{}}
What are the offloading tools and techniques \\
used to orchestrate the microservices in the \\
Edge Computing Scenario?\end{tabular}                                                                                                                          & \multicolumn{1}{c}{RQ 4}                 \\  \midrule
11          & Challenges/Problems                                                                       & \begin{tabular}[c]{@{}l@{}}What challenges/problems are recognized in the literature\\ relating $\mu$Services and Edge Computing?\end{tabular}                                                                                                                                                         & \multicolumn{1}{c}{RQ 5}                 \\ \bottomrule
\end{tabular}\caption{Extraction Form}
\label{tab:extractionForm}
\end{table*}

\subsection{Reporting - Overview of Selected Studies}
\label{subsubsec:overview}

\begin{figure}[h]
        \centering
         \fbox{\includegraphics[width=0.7\textwidth]{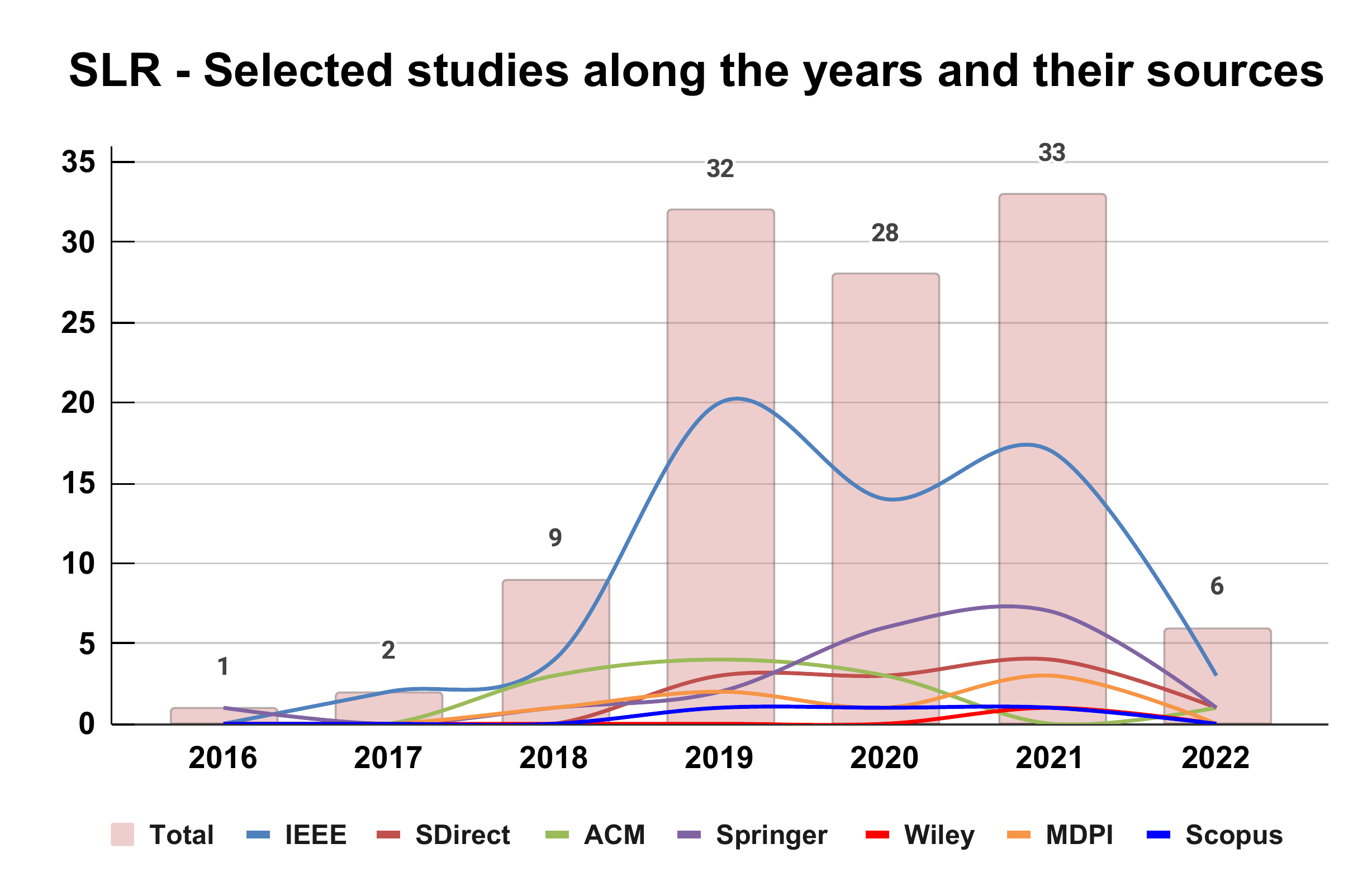}}
        \caption{Temporal perspective of the selected studies}
          \label{fig:slr_temporal}
   \end{figure}

All 111 selected studies appeared between 2014 and 2022\footnote{End of May of 2022}. Figure \ref{fig:slr_temporal} shows a temporal perspective of the papers classified by their databases. As we noticed, there has been an increase in published studies after 2018. It occurred due to both technologies becoming mature in the academy and industry, as pointed out by a Gartner report\cite{gartnerreportedge_blog,gartnerhypecycleedge2021}.

    \begin{figure*}[h!]
        \begin{minipage}{.49\textwidth}
          \fbox{\includegraphics[width=0.97\textwidth]{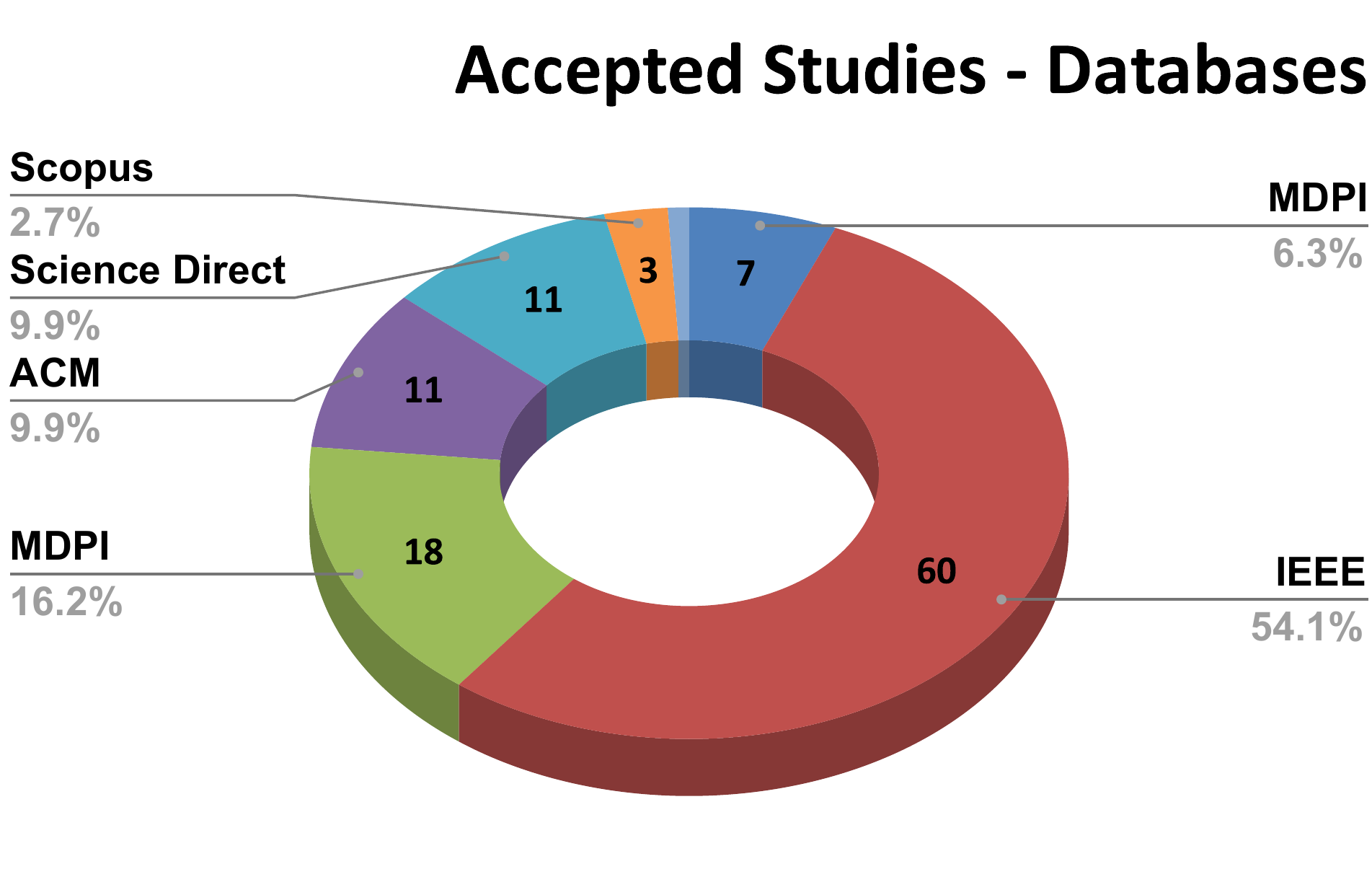}}
          \captionof{figure}{Select studies per Database}
          \label{fig:slr_databases}
        \end{minipage}
        \begin{minipage}{.49\textwidth}

            \flushright
            \small
            \begin{tabular}{@{}lccc@{}}
            \toprule
            \textbf{Database}                                                   & \multicolumn{1}{l}{\textbf{Studies}}                   & \textbf{\#} & \textbf{\%} \\ \midrule
            \begin{tabular}[c]{@{}l@{}}IEEE\end{tabular}         & \begin{tabular}[c]{@{}c@{}}
            
            \cite{paper_58,paper_41,paper_24,paper_26,paper_10,paper_78,paper_27}\\
            \cite{paper_38,paper_95,paper_87,paper_22,paper_16,paper_94}\\            \cite{paper_104,paper_80,paper_84,paper_61,paper_6,paper_31,paper_39,paper_2,paper_7,paper_9,paper_11,paper_12,paper_13,paper_14,paper_15,paper_20,paper_23,paper_30,paper_48,paper_51,paper_53,paper_54,paper_55,paper_60,paper_68,paper_70,paper_71,paper_72,paper_73,paper_75,paper_79,paper_82,paper_83,paper_85,paper_86,paper_88,paper_89,paper_90,paper_91,paper_92,paper_93,paper_96,paper_97,paper_99,paper_105,paper_106,paper_108}

            \end{tabular}          & 60          & 54.1        \\ \midrule
            \begin{tabular}[c]{@{}l@{}}Springer\end{tabular}   & \begin{tabular}[c]{@{}c@{}}
            
            \cite{paper_98,paper_102,paper_46,paper_65,paper_4,paper_17,paper_28,paper_40,paper_42,paper_47,paper_49,paper_50,paper_57,paper_63,paper_69,paper_77,paper_100,paper_101}
            
            \end{tabular} & 18          & 16.2        \\ \midrule
            \begin{tabular}[c]{@{}l@{}}Elsevier\end{tabular} & \begin{tabular}[c]{@{}c@{}}
            \cite{paper_103,paper_5,paper_18,paper_33,paper_35,paper_43,paper_56,paper_59,paper_62,paper_66,paper_110}
            
            \end{tabular}          & 11          & 9.9         \\ \midrule
            ACM    & \begin{tabular}[c]{@{}c@{}}
            \cite{paper_45,paper_19,paper_32,paper_34,paper_36,paper_44,paper_52,paper_64,paper_74,paper_81,paper_109}
            \end{tabular} & 11 & 9.9 \\ \midrule
            MDPI   & \begin{tabular}[c]{@{}c@{}}\cite{paper_8,paper_1,paper_0,paper_3,paper_25,paper_29,paper_107}\end{tabular} & 7  & 6.3 \\ \midrule
            Scopus & \cite{paper_21,paper_67,paper_76}                                            & 3  & 2.7 \\ \midrule
            Wiley  & \cite{paper_37}                                             & 1  & 0.9 \\ \bottomrule
            \end{tabular}
            \captionof{table}{Studies for each Database}

        \end{minipage}
\end{figure*}


Moreover, Figure \ref{fig:slr_databases} illustrates the studies categorized by their respective databases. The Scopus database indexes 32 selected titles, IEEEXplore indexes 17, 14 in Springer, 8 in ScienceDirect and Web of Science, and 4 in ACM.

\begin{figure*}[h]
    \begin{minipage}{.495\textwidth}
        \centering
        \fbox{\includegraphics[width=0.9\textwidth]{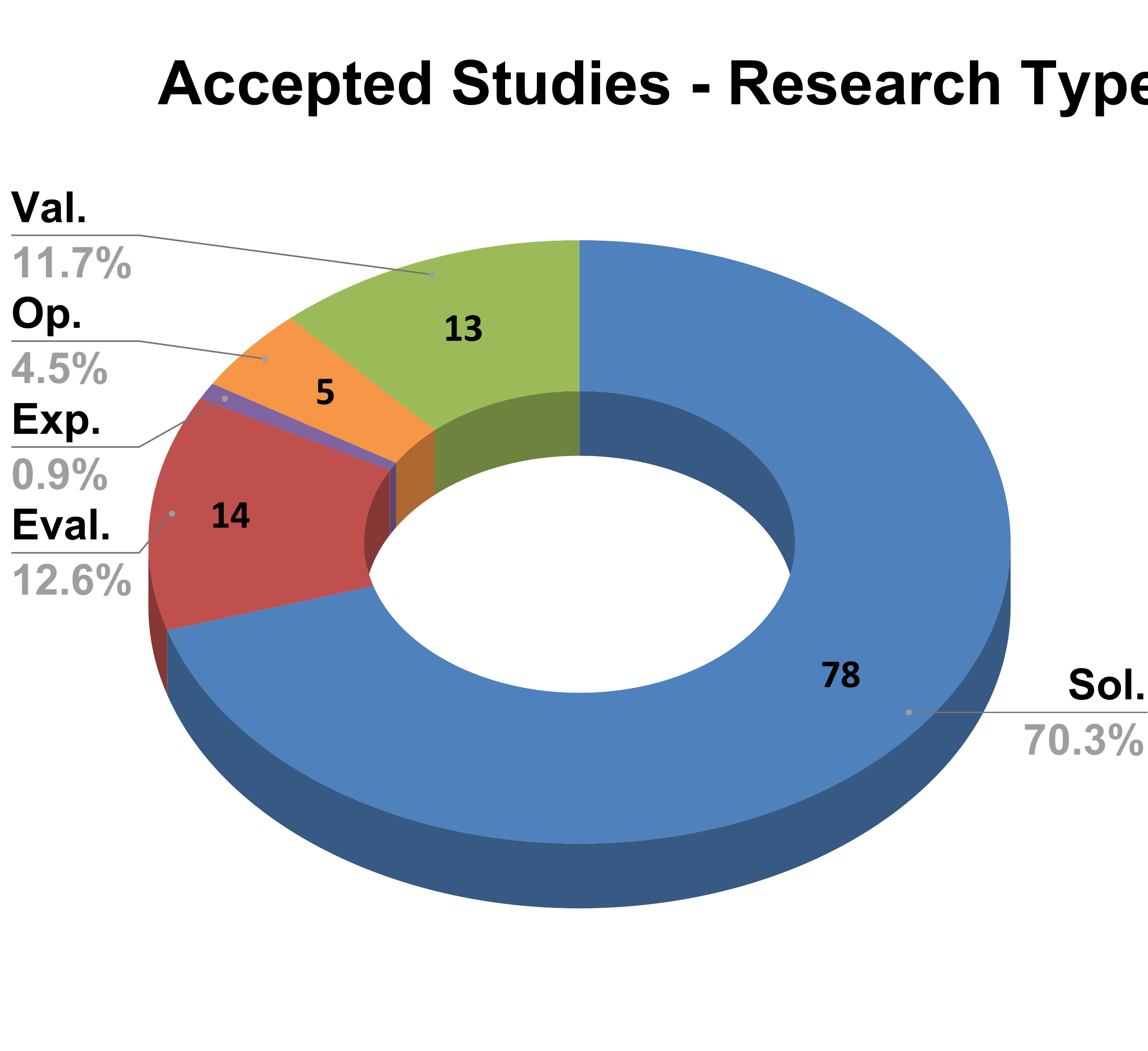}}
        \captionof{figure}{Research type - Studies Overview}
        \label{fig:slr_researchtype}
    \end{minipage}
    \begin{minipage}{.495\textwidth}
        \flushright
        \small
        \begin{tabular}{@{}llcc@{}}
            \toprule
            \textbf{Res. Type} & \textbf{Studies}                                           & \textbf{\#} & \textbf{\%} \\ \midrule
            \begin{tabular}[c]{@{}l@{}}Solution \\ Proposal \end{tabular}      & \begin{tabular}[c]{@{}l@{}}
          
            \cite{paper_58,paper_41,paper_98,paper_24,paper_26,paper_10,paper_78,paper_103,paper_27,paper_38,paper_95,paper_102,paper_87,paper_8,paper_16}\\
            \cite{paper_94,paper_104,paper_65,paper_80,paper_84,paper_61,paper_1,paper_31,paper_39,paper_9,paper_12}\\ 
            \cite{paper_15,paper_20,paper_23,paper_30,paper_48,paper_51,paper_54,paper_55,paper_60,paper_68,paper_71,paper_72,paper_73,paper_82}\\
            \cite{paper_83,paper_86,paper_88,paper_90,paper_91,paper_93,paper_96,paper_97,paper_99,paper_105,paper_106,paper_108,paper_4,paper_17}\\
            
            \cite{paper_50,paper_63,paper_69,paper_77,paper_101,paper_5,paper_43}\\
            
            \cite{paper_110,paper_32,paper_34,paper_52,paper_64,paper_74,paper_81,paper_109,paper_0,paper_3,paper_25,paper_29,paper_107,paper_21,paper_67,paper_76,paper_37}\\

            \end{tabular}          & 78             & 70.3        \\ \midrule
            \begin{tabular}[c]{@{}l@{}}Evaluation \\Research  \end{tabular}   & \begin{tabular}[c]{@{}l@{}}
            
            \cite{paper_46,paper_6,paper_2,paper_14,paper_53,paper_70}\\
            \cite{paper_28,paper_40,paper_42,paper_49,paper_33}\\
            \cite{paper_35,paper_56,paper_62}

            \end{tabular}          & 14             & 13.5           \\ \midrule
            \begin{tabular}[c]{@{}l@{}}Validation \\Research  \end{tabular}   & \begin{tabular}[c]{@{}l@{}}
            
            \cite{paper_22,paper_13,paper_75,paper_79,paper_85,paper_89}\\
            \cite{paper_92,paper_47,paper_100,paper_18}\\
            \cite{paper_59,paper_36,paper_44}
            
            \end{tabular} & 13              & 10.8           \\ \midrule
            \begin{tabular}[c]{@{}l@{}}Opinion    \end{tabular}      & \begin{tabular}[c]{@{}l@{}}
            \cite{paper_45,paper_11,paper_57,paper_66,paper_19}
            \end{tabular}          & 5              & 4.5           \\ \midrule
            \begin{tabular}[c]{@{}l@{}}Exp. Report  \end{tabular}     & \begin{tabular}[c]{@{}l@{}}\cite{paper_7}\end{tabular}              & 1              & 0.9           \\ \bottomrule
        \end{tabular}
        \captionof{table}{Research Type - Listing Studies}
        
    \end{minipage}
\end{figure*}

We also performed a type classification using these categories: Book Chapter, Conference Paper, and Journal Article. We consider Book Chapters different from Journal Articles, especially in the case of Springer books. Figure \ref{fig:slr_researchtype} shows this classification. Most select studies are from Conference Papers (44 titles), followed by Journal Articles with 37 entries and Book Chapters with 4 entries. 

\begin{figure*}[h]
    \begin{minipage}{.495\textwidth}
        \centering
          \fbox{\includegraphics[width=0.9\textwidth]{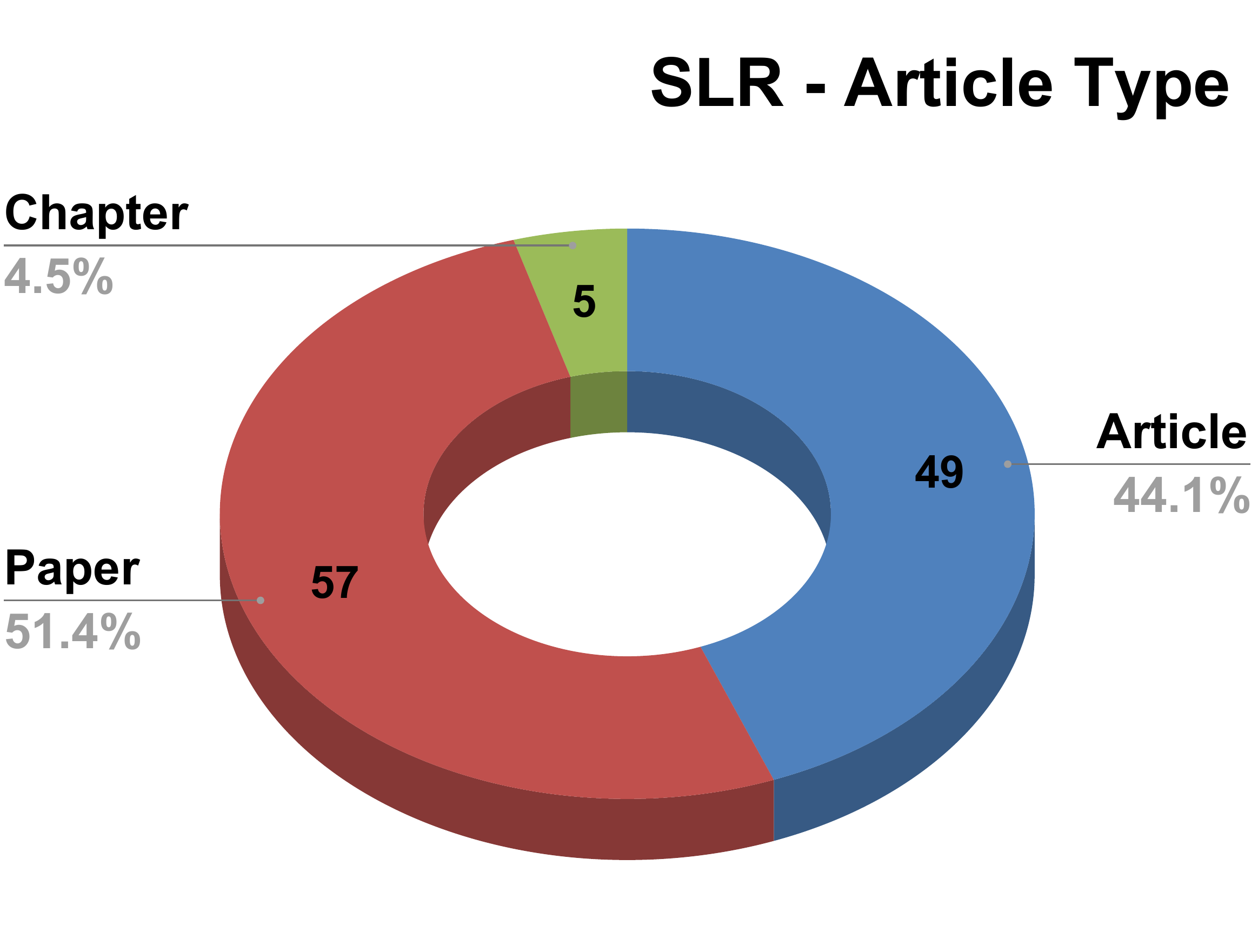}}
          \captionof{figure}{Article type - Overview Studies}
          \label{fig:slr_articletype}
    \end{minipage}
    \begin{minipage}{.495\textwidth}
        \small
        \centering
        \begin{tabular}{@{}llcc@{}}
\toprule
\textbf{Art. Type} & \textbf{Studies}                                           & \textbf{\#} & \textbf{\%} \\ \midrule
\begin{tabular}[c]{@{}l@{}}Conference \\ Paper \end{tabular}      & 
\begin{tabular}[c]{@{}l@{}}
\cite{paper_41,paper_24,paper_26,paper_10,paper_27,paper_22,paper_16}\\
\cite{paper_104,paper_80,paper_61,paper_6,paper_45,paper_31,paper_39}\\
\cite{paper_9,paper_11,paper_12,paper_13,paper_14,paper_15,paper_23,paper_30,paper_48,paper_51,paper_54,paper_60,paper_68}\\
\cite{paper_72,paper_73,paper_75,paper_79,paper_82,paper_83,paper_85,paper_86,paper_88,paper_89,paper_90,paper_91,paper_92,paper_96,paper_105,paper_106}\\
\cite{paper_5,paper_19,paper_32,paper_34,paper_36,paper_44,paper_52,paper_64,paper_74,paper_81,paper_109,paper_21,paper_76}\\

\end{tabular}          & 57             & 51.4        \\ \midrule
\begin{tabular}[c]{@{}l@{}}Journal\\ Article \end{tabular}      & \begin{tabular}[c]{@{}l@{}}

\cite{paper_58,paper_98,paper_78,paper_103,paper_38,paper_95,paper_46}\\
\cite{paper_87,paper_8,paper_94,paper_84,paper_1,paper_2,paper_7}\\
\cite{paper_20,paper_53,paper_55,paper_70,paper_71,paper_93}\\
\cite{paper_97,paper_99,paper_108,paper_4}\\
\cite{paper_28,paper_40,paper_42,paper_50,paper_57,paper_63,paper_69,paper_77,paper_100,paper_101}\\
\cite{paper_18,paper_33,paper_35,paper_43,paper_56,paper_59,paper_62,paper_66,paper_110, paper_0,paper_3,paper_25,paper_29,paper_107,paper_67,paper_37}

\textbf{}
\end{tabular} & 49              & 44.1          \\ \midrule
\begin{tabular}[c]{@{}l@{}}Book Chapter \end{tabular}         & \begin{tabular}[c]{@{}l@{}}\cite{paper_102,paper_65,paper_17,paper_47,paper_49}\end{tabular}          & 5              & 4.5           \\ \bottomrule
\end{tabular}
        \captionof{table}{Article Type - Listing Studies}
        
    \end{minipage}
\end{figure*}


Further, we classified studies by their category: Solution Proposal, Validation Research, Evaluation Research, Experience Report, or Opinion Paper. Figure \ref{fig:slr_researchtype} presents the study categories, and Table 7 lists the studies by their category. Most of the studies are Solution Proposals (58 titles), highlighting how Microservices and Edge Computing have been used to propose solutions to the industry and the academy. Moreover, the Evaluation Research papers (14 titles) demonstrated how the projects have matured. Finally, the Opinion Papers (5 titles) discuss how distributed applications (e.g., based on microservices) could face the Edge Computing scenario \cite{paper_57,paper_66}. Also, the authors discuss the implications of using Edge Computing as the predominant paradigm in terms of security and privacy \cite{paper_66}, latency awareness \cite{paper_19}, design patterns \cite{paper_45}, and other distributed systems issues \cite{paper_11,paper_57}.
    
\begin{figure*}[h!]
    
    \begin{minipage}{.49\textwidth}
        \vspace{.7cm}
        \centering
        \fbox{\includegraphics[trim={1.5cm 0 0 0},width=.95\textwidth]{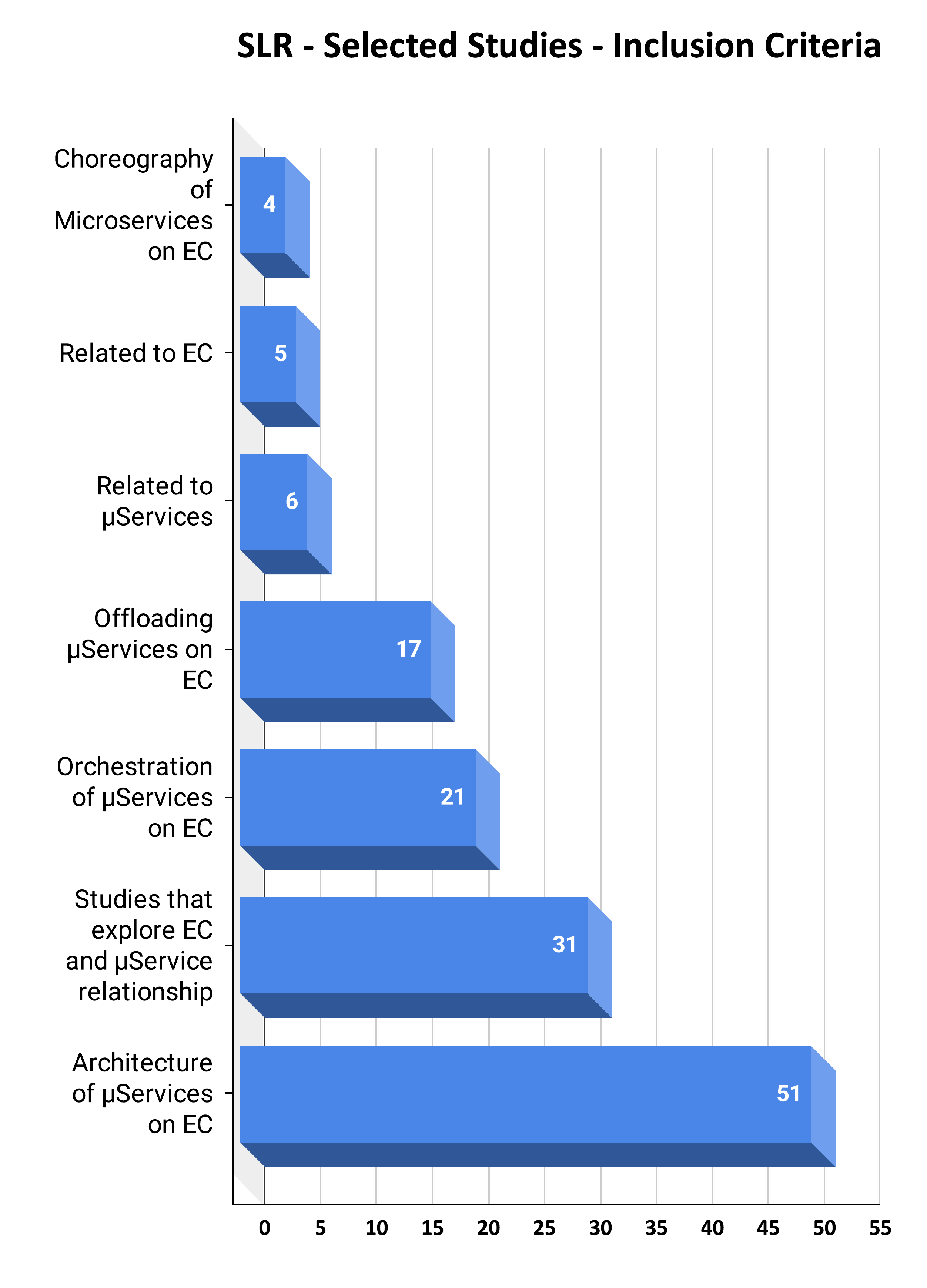}}
        \caption{Inclusion Criteria - Studies Overview}
        \label{fig:acceptedStudies_criteria}

    \end{minipage}
    \begin{minipage}{.49\textwidth}
    \small
    \flushright
    \captionof*{table}{$\alpha$ - \scriptsize The total amount is greater than 111 due to some selected studies belong to more than 1 inclusion criteria}
    \vspace{-.3cm}
    \begin{tabular}{@{}llcc@{}}
    
\toprule
\textbf{Incl. Criteria}                                                 & \textbf{Studies}                                       & \footnotesize\textbf{$\#^\alpha$} & \footnotesize\textbf{\%} \\ \midrule
\begin{tabular}[c]{@{}l@{}}1) Rel. $\mu$Services\\ and EC\end{tabular}  & \begin{tabular}[c]{@{}l@{}}

\cite{paper_41,paper_27,paper_46,paper_65,paper_84,paper_11}\\
\cite{paper_12,paper_51,paper_55,paper_60,paper_68,paper_72,paper_75}\\

\cite{paper_82,paper_89,paper_4,paper_17,paper_28}\\
\cite{paper_47,paper_49,paper_50,paper_63,paper_77}\\
\cite{paper_35,paper_43,paper_110,paper_34}\\
\cite{paper_52,paper_81,paper_0,paper_3}

\end{tabular}          & 31              & 27.9           \\ \midrule
\begin{tabular}[c]{@{}l@{}}2) $\mu$Services \\ Arch. on EC\end{tabular} & \begin{tabular}[c]{@{}l@{}}

\cite{paper_98,paper_26,paper_78,paper_103,paper_95,paper_102}\\
\cite{paper_87,paper_8,paper_16,paper_94,paper_104,paper_80}\\ 
\cite{paper_45,paper_39,paper_7,paper_14,paper_15,paper_20,paper_23,paper_48}\\
\cite{paper_53,paper_54,paper_70,paper_71,paper_73,paper_83}\\
\cite{paper_86,paper_88,paper_90,paper_91,paper_92,paper_93,paper_96,paper_97,paper_99,paper_105,paper_106,paper_108,paper_100}\\
\cite{paper_5,paper_62,paper_19,paper_36}\\
\cite{paper_44,paper_74,paper_25,paper_107,paper_21,paper_67,paper_76,paper_37}


\end{tabular} & 51             & 46           \\ \midrule
\begin{tabular}[c]{@{}l@{}}3) Orchestrating \\ $\mu$Services on EC\end{tabular} & \begin{tabular}[c]{@{}l@{}}

\cite{paper_58,paper_98,paper_24,paper_38,paper_95,paper_102,paper_87}\\
\cite{paper_22,paper_94,paper_1,paper_9,paper_13,paper_20}\\
\cite{paper_85,paper_88,paper_108,paper_57,paper_101}\\
\cite{paper_56,paper_66,paper_29}

\end{tabular} & 21 & 18.9 \\ \midrule
\begin{tabular}[c]{@{}l@{}}4) Choreography \\ $\mu$Services on EC\end{tabular}  & \begin{tabular}[c]{@{}l@{}}\cite{paper_10,paper_87,paper_2,paper_85}\end{tabular} & 4 & 3.6 \\ \midrule

\begin{tabular}[c]{@{}l@{}}5) Offloading \\ $\mu$Services on EC\end{tabular}    & \begin{tabular}[c]{@{}l@{}} 

\cite{paper_95,paper_87,paper_22,paper_16,paper_61,paper_6}\\
\cite{paper_1,paper_9,paper_13,paper_14,paper_85,paper_93}\\
\cite{paper_106,paper_108,paper_52,paper_25,paper_109} \\

\end{tabular}

& 17 & 15.3 \\ \midrule
\begin{tabular}[c]{@{}l@{}}6) Related to \\ $\mu$Services\end{tabular}  & \begin{tabular}[c]{@{}l@{}}\cite{paper_31,paper_30,paper_79}\\\cite{paper_40,paper_18,paper_64}\end{tabular} & 6                    & 5.4           \\ \midrule
7) Related to EC                                                        & \begin{tabular}[c]{@{}l@{}}\cite{paper_42,paper_69,paper_33,paper_59,paper_32}\end{tabular} & 5              & 4.5          \\ \bottomrule
\end{tabular}
\captionof{table}{Inclusion Criteria - Listing Studies}
    \end{minipage}
\end{figure*}

Finally, we classified the studies by their inclusion criteria. Figure \ref{fig:acceptedStudies_criteria} demonstrates those criteria and Table \ref{tab:inclusionCriteria} lists the studies.

\begin{table}[]
\centering

\caption{Selected Studies - Inclusion Criteria}
\label{tab:inclusionCriteria}
\end{table}

\subsection{The Systematic Review - Threats to validity}

We employ four threats of validity presented in \cite{wohlin2012experimentationSLR} and in \cite{kitchenham2015evidence}: construct, internal, conclusion, and external. SLR guidelines strongly recommend these threats to avoid research bias \cite{kitchenham2007slr}. One important point to handle those threats is the care in the design and conduction of the review. Since the first author is a Ph.D. candidate, the search string design, protocol, and preliminary results have been analyzed and discussed with a board of specialists during his qualification exam, helping to mitigate the threats. Experienced researchers in Software Engineering, Edge Computing, and Distributed Architectures formed this board.

\begin{itemize}

\item Construct validity is related to the design and construction of the protocol. Moreover, how the drawn conclusions are related to the theory supporting the study. We built the protocol using well-established databases to minimize the threats to this validity. The research string was designed under well-established academic terms [\cite{edge_a_prime_2018,liu_survey_2019,slr_vilela2017integration,petursdottir2018applyingSLRThreats}  and industry \cite{gartner_edge_report,gartnerhypecycleedge2021,gartnerreportedge_blog}.

\item  Internal validity concerns how the outcomes of the SLR are affected by a factor without the researchers’ knowledge \cite{kitchenham2015evidence}. In other words, there might be wrong conclusions due to causal issues between the theory (protocol) and the results \cite{wieringa2006requirementsSLR}. To avoid this threat, we choose the SLR methodology to review the literature due to its interactive process. 

\item Conclusion validity reflects how the authors analyzed the outcomes of the current study and how it was conducted \cite{kitchenham2015evidence}. Also, how do the researchers draw the correct conclusion about the treatment and the results \cite{wohlin2012experimentationSLR}? To steer clear of this threat, we carefully built our research string using logic and searching operands to maximize the number of primary studies selected. We collected 1818 primary studies to analyze, covering the 2014-2022 period in this selection.

\item External validity is related to how the conclusions of a study may be generalized to an interested population \cite{kitchenham2015evidence}. Applying it to an SLR is the condition that limits the ability to generalize the SLR results \cite{wohlin2012experimentationSLR} to verify possible research gaps, trends, and opportunities. To mitigate this threat, we started the search process after several attempts, whereas the authors achieved a consensus.
        
\end{itemize}

\begin{figure}[h!]
    \centering
    \fbox{\includegraphics[width=0.75\textwidth]{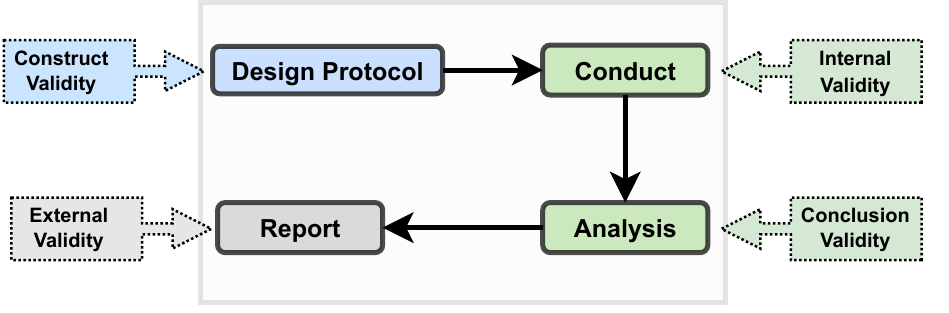}}
    \caption{Threats to a SLR validity and where they appear - adapted from \cite{kitchenham2015evidence}}
   	\label{fig:slr_threats}
\end{figure}

\section{The Systematic Review - Results and Discussion}
\label{sec:slr_results}

\begin{sidewaysfigure}
    \fbox{\includegraphics[width=.95\textwidth]{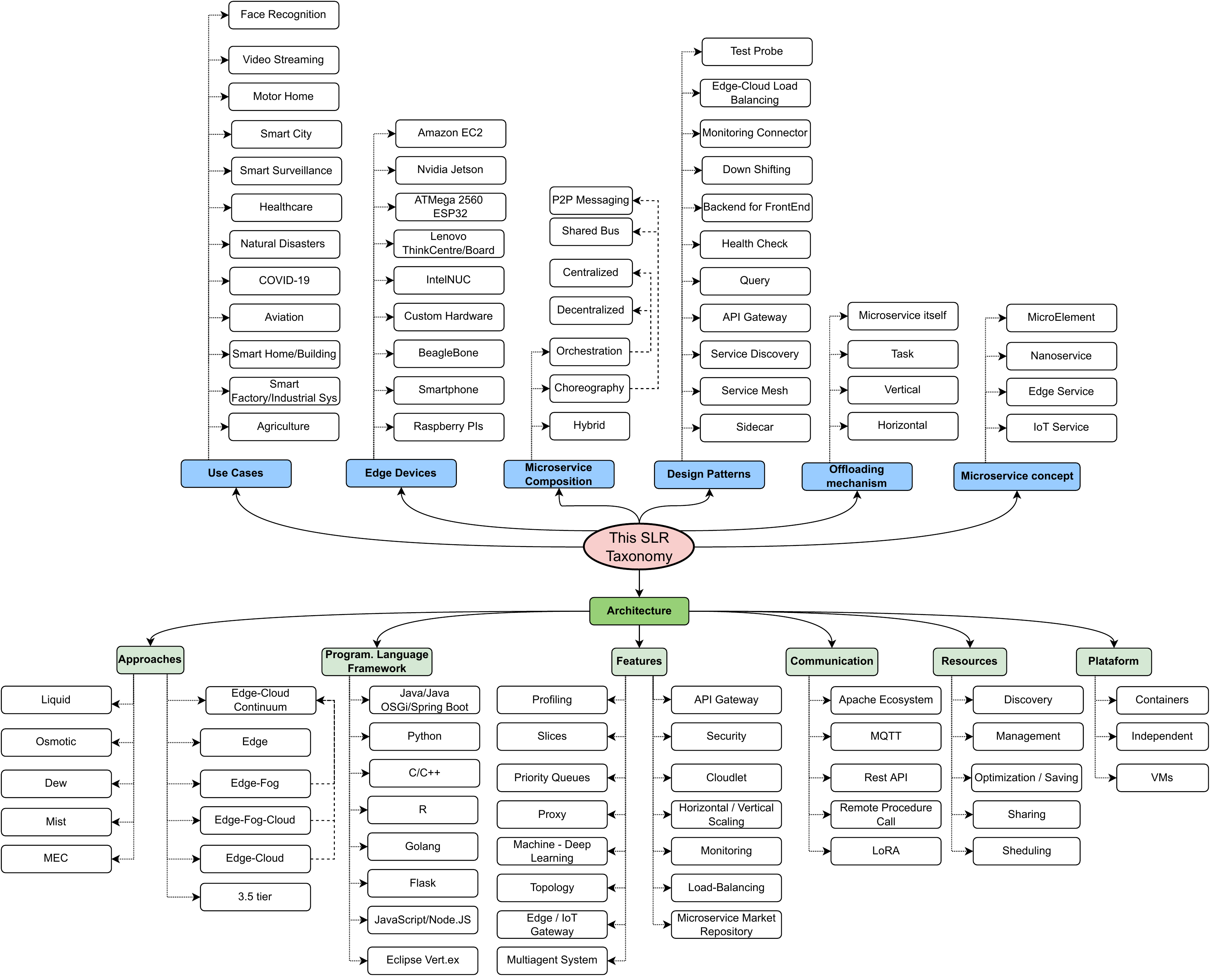}}
    \caption{SLR - Taxonomy}
    \label{fig:slr_taxonomy}
\end{sidewaysfigure}

In this section, we discuss the answers to the seven Research Questions (described in Table \ref{tab:researchQuestions}). We classified the selected studies in a taxonomy according to their categories based on the research string keywords. Initially, we arranged them in terms of architectural approaches and features, composition, and offloading. However, whereas the SLR went through, we also classified the studies into additional categories, such as use cases, edge devices, design patterns, programming languages/frameworks, communication, resources, and platforms. Indeed, these categories are nested in the research string keywords. We used them to bring a broader and deeper perspective of the selected studies and their contributions. Figure \ref{fig:slr_taxonomy} presents the taxonomy.

\subsection{RQ1 - What are the existing microservice architectures' approaches and features to Edge Computing?}
\label{subsec:rq4}

This question aims to identify how different architecture features and approaches have employed Microservice architectures in Edge Computing. This question leads us to two sub-questions. (RQ 1.1) The first focuses on the advantages and disadvantages of utilizing microservices in Edge Computing. (RQ 1.2) The second addresses which microservice characteristics take advantage of running in Edge Computing. If so, how essential are these to run in the Edge? 

We define an architecture approach in terms of how many and on which tiers the architecture work. We refer to tier as a layer in the architecture. Most selected studies consider Edge, Fog, and Cloud as the standard tiers \cite{buyya_fog_2018,buyya2019fogedgebook,edge_a_prime_2018}. A single-tier approach will work only in Edge. On the other hand, n-tier approaches (2-tier and 3-tier) will work in Edge+Fog, Edge+Cloud, or even Edge + Fog + Cloud. 

\begin{figure*}[h!]
    \begin{minipage}{.49\textwidth}
        \vspace{1.6cm}
        \centering
        \fbox{\includegraphics[trim={.25cm .2cm .5cm 1cm},width=.95\textwidth,height=11.45cm]{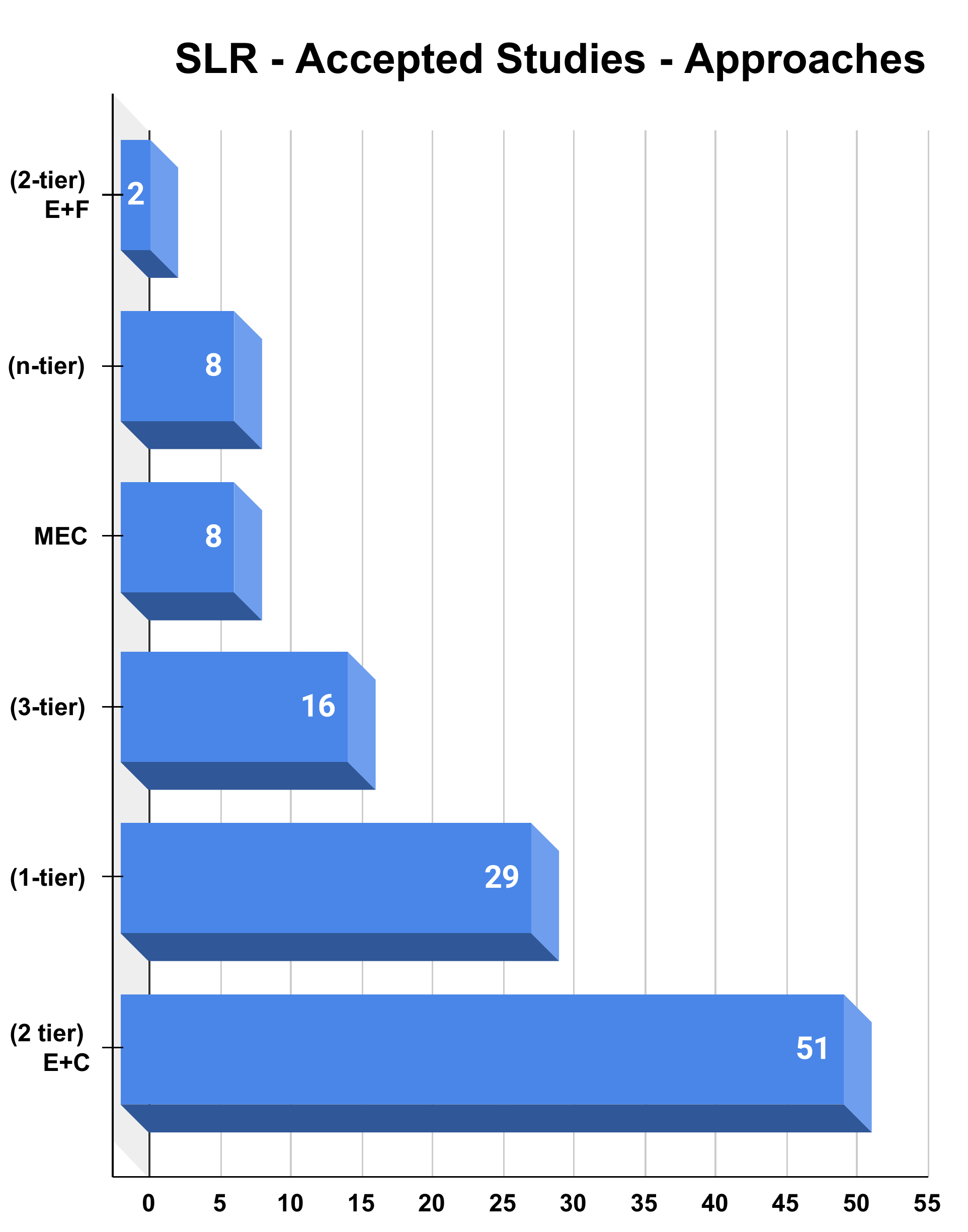}}
        \caption{Architecture Approaches - Studies Overview}
        \label{fig:slr_approaches}

    \end{minipage}
    \begin{minipage}{.49\textwidth}
\small
\flushright
\captionof*{table}{$\beta$ - \scriptsize There are two papers (\cite{paper_31,paper_2}) classified as \textit{not applicable} due to the studies do not focus in any specific approach.}
\vspace{-.3cm}
\captionof*{table}{$\gamma$ - \scriptsize The total amount is greater than 111 due four papers were classified into two approaches. 1) \cite{paper_102} is 3-tier and Continuum; 2) \cite{paper_103} is 3-tier and 3.5 tier; 3) \cite{paper_20,paper_27} are 2-tier (E+C) and Mist}
\vspace{-.3cm}

\begin{tabular}{@{}llcc@{}}
\toprule
\textbf{Approach$^\beta$}             & \textbf{Reference} & \footnotesize\textbf{$\#^\gamma$} & \footnotesize\textbf{\%} \\ \midrule
(1-tier) Edge & 
\begin{tabular}[c]{@{}l@{}}

\cite{paper_24,paper_26,paper_38,paper_94,paper_61}\\
\cite{paper_6,paper_30,paper_54,paper_60,paper_75}\\ 
\cite{paper_70,paper_71,paper_72,paper_79,paper_90,paper_91}\\
\cite{paper_97,paper_99,paper_40,paper_63}\\
\cite{paper_69,paper_35,paper_56,paper_36}\\
\cite{paper_64,paper_74,paper_0}\\
\cite{paper_76,paper_37}
\end{tabular} & 29 & 25.5
\\ \hline
\begin{tabular}[c]{@{}l@{}}(2-tier) Edge + \\ Cloud (E+C)   \end{tabular}             &  \begin{tabular}[c]{@{}l@{}}

\cite{paper_10,paper_78,paper_27,paper_95,paper_46}\\
\cite{paper_22,paper_8,paper_16,paper_104,paper_65}\\
\cite{paper_45,paper_39,paper_7,paper_12,paper_13}\\
\cite{paper_20,paper_23,paper_48,paper_55,paper_68}\\
\cite{paper_73,paper_82,paper_83,paper_86,paper_88,paper_89}\\
\cite{paper_92,paper_93,paper_96,paper_108,paper_17} \\
\cite{paper_28,paper_42,paper_47,paper_49,paper_50,paper_77}\\
\cite{paper_101,paper_5,paper_43,paper_62}\\

\cite{paper_110,paper_19,paper_32,paper_34,paper_81,paper_109}\\
\cite{paper_3,paper_25,paper_29,paper_107,paper_67}\\

\end{tabular}    & 51 & 44.7       \\ \hline
\begin{tabular}[c]{@{}l@{}}(2-tier) Edge + Fog \\ (E+F) \end{tabular}                  & \cite{paper_11,paper_21}  & 2 & 1.8             \\ \hline
(2-tier) MEC                         & 
\begin{tabular}[c]{@{}l@{}}
\cite{paper_41,paper_14,paper_84,paper_85,paper_105}\\
\cite{paper_106,paper_4,paper_100}    
\end{tabular}
& 8 & 7\\ 
\hline
\begin{tabular}[c]{@{}l@{}}(3-tier) Edge + \\ Fog + Cloud      \end{tabular}    &
\begin{tabular}[c]{@{}l@{}}

\cite{paper_58,paper_98,paper_102,paper_103,paper_87}\\
\cite{paper_1,paper_9,paper_15,paper_51,paper_53}\\
\cite{paper_57,paper_18,paper_33}\\
\cite{paper_66,paper_52,paper_21}\end{tabular}       & 16 & 14          \\ \hline

\begin{tabular}[c]{@{}l@{}}
(n-tier) Edge + \footnotesize(Dew, \\ \footnotesize(Mist, 3.5, Osmotic,\\
\footnotesize Liquid, Continuum) \normalsize \end{tabular}& \begin{tabular}[c]{@{}l@{}}\cite{paper_103,paper_27,paper_95,paper_102}\\
\cite{paper_80,paper_20,paper_59} \end{tabular} & 8 & 7\\
\bottomrule
\end{tabular}
\captionof{table}{Architecture Approaches - Listing Studies}
\label{tab:slr_approaches}

\end{minipage}
\end{figure*}

Beyond this standard, some authors have been proposing intermediary tiers above and below the Edge. Firstly, the research in \cite{paper_103,paper_107} introduces a 3.5 tier. The accessory ".5" tier lies between the Edge (Application Layer) and the IoT (Perception Layer). In this tier, microservices powered by a deep-learning technique aim to explore and improve the load-balancing in the Edge nodes above. Secondly, Dew and Mist Computing are targeting IoT devices, working towards the "extreme" endpoints of the Edge tier. Therefore, they employ the Edge tier as an auxiliary layer to their Dew/Mist Computing  \cite{paper_27}. Thirdly, Osmotic Computing proposes the Microelement concept, which introduces a single structure to encapsulate the microservice and its data. Thus, this microelement flows between Edge and Cloud domains, making an osmosis movement ~\cite{paper_20}. Based on this "flow" concept, some authors employ the \textit{EdgeCloud Continuum} approach, where the Edge and Cloud are distinct tiers. However, the microservices and IoT services see them as the same tier. Likewise, Liquid Computing ~\cite{paper_107} proposes something similar, with no noticeable border between the tiers. Thus, the microservices move around the tiers according to user/system requirements to achieve resource savings concurrently with QoS satisfaction. Table \ref{tab:slr_approaches} and Figure \ref{fig:slr_approaches} summarize and illustrate the selected studies by their respective approaches.

Some studies are exploring a higher concept that works towards and along with tiers: the architecture topology \cite{paper_87,paper_95}. Topologies like master-slave, mesh, client-server, hierarchical, and flat were employed to dispose of and manage the edge nodes and microservice instances. One can improve resource allocation and guarantee the overall QoS by choosing a proper topology~\cite{paper_95}. For example, if the microservices lay in a mesh topology, they can provide an abstract layer between the IoT and the Cloud. This layer could pre-process, filter, analyze, and annotate the sensor data (IoT) and then send them to Cloud \cite{paper_86,paper_92}. Moreover, in \cite{paper_85}, the authors explain how the microservice mesh can benefit resource sharing. 

 Regarding architecture features, we define them as every tool, platform, and pattern that bases the architecture itself. To guide the following discussion, we list all the features employed and suggested by the 111 selected studies in Table \ref{tab:slr_features}. 

\begin{table}[h]
\centering
\small
\begin{tabular}{@{}l|l@{}}
\toprule
\textbf{Architecture Features}          & \textbf{Reference}                                                           \\ \midrule

API Gateway                              & \begin{tabular}[c]{@{}l@{}}

\cite{paper_78,paper_22,paper_7,paper_15,paper_20,paper_30,paper_48,paper_71,paper_82,paper_83,paper_85,paper_88,paper_90,paper_17,paper_45,paper_0,paper_21,paper_76}

\end{tabular}  \\ \midrule
Blockchain                               & \cite{paper_58,paper_13,paper_15,paper_48,paper_97,paper_0,paper_21}                               \\ \midrule
Cloudlet                                 & \cite{paper_16}                                                            \\ \midrule
Smart/Edge Gateway                             & \begin{tabular}[c]{@{}l@{}}

\cite{paper_10,paper_70,paper_84,paper_93,paper_40,paper_77,paper_5,paper_36,paper_3,paper_25,paper_107,paper_67} 

\end{tabular}                     \\ \midrule
LoRA/CoAP Message Protocols                        &

\cite{paper_27,paper_87,paper_94,paper_18,paper_34}     

\\ \midrule
Database                    & \begin{tabular}[c]{@{}l@{}}

\cite{paper_10,paper_78,paper_27,paper_22,paper_6,paper_20,paper_23,paper_48,paper_71,paper_82,paper_83,paper_77,paper_74,paper_76}

\end{tabular} \\ \midrule
\begin{tabular}[c]{@{}l@{}}Monitoring Agent/Server  \\ Timeseries DB \end{tabular}& \begin{tabular}[c]{@{}l@{}}
\cite{paper_1,paper_12,paper_82,paper_88,paper_93,paper_17,paper_47,paper_50,paper_67,paper_34}                 \end{tabular}                        \\ \midrule
MQTT Message Protocol & 
\begin{tabular}[c]{@{}l@{}}
\cite{paper_98,paper_27,paper_87,paper_8,paper_60,paper_93,paper_28,paper_40,paper_42,paper_77,paper_34,paper_67,paper_37}

\end{tabular}\\ \midrule
Multiagent System & \cite{paper_75,paper_50,paper_59} \\ \midrule
                                                            
Priority Queues                          & \cite{paper_14,paper_60}
\\\midrule
Profiling & \cite{paper_94,paper_91} \\ \midrule
Proxy & \cite{paper_103,paper_51,paper_73,paper_82,paper_32,paper_34,paper_67}
\\\bottomrule
\end{tabular}
\caption{Architecture Features}
\label{tab:slr_features}
\end{table}



Several research projects employ the API Gateway feature, providing an entry point to the whole microservice architecture ~\cite{paper_0}. The gateway could be designed and implemented as a standalone microservice~\cite{microservicespatternsBook}. Its most common use case is for users requesting functionalities provided by the system through an API Gateway~\cite{paper_7}. However, API Gateways work in a centralized mode, leading to a system bottleneck. A multi-purpose API Gateway is introduced in \cite{paper_30}, employing textit{n}-API Gateways according to the users' demands and interests. Multiple API Gateways enable the system to balance user requests, leading to a load balance at the API Gateway level.

Edge Gateways are related to the IoT devices in an Edge Computing scenario by providing connection among them and between them and the Edge Servers \cite{paper_46} or even to the Cloud. Edge Gateways can also supply several services to the IoT devices, such as protocol conversion, device management, storage, local processing, edge-cloud integration \cite{paper_10}, and security and access control \cite{paper_53}. 

Some microservice architectures handle security aspects. The API Gateway and the Edge Gateway can be essential in access control\cite{paper_51,paper_82}. For example, the Edge Gateway can validate payloads, providing a security layer to IoT devices. Some authors tackled security issues at the transaction level, like in the IBM Hyperledge Fabric, applying blockchain technology to grant access to the system and control the transactions among microservice peers~\cite{paper_0}. Blockchain was also used to assure transaction control in an n-tier microservice architecture ~\cite{paper_15,paper_21}, where the hash function is calculated in the Fog or even in the Cloud. Still, regarding security, the authors in \cite{paper_16} provide a Cloudlet\footnote{A edge node working as a mini-cloud-datacenter in the edge \cite{satyanarayanan2009case}} node to separate the edge devices from the external network, acting similarly to a proxy service \cite{paper_32}. 

Resource availability through proper management and optimization is key to achieving a satisfactory QoS in the Edge Computing scenario. It is also an issue in the Cloud but is urgent in the Edge. Some architectural approaches have been targeting this issue. For instance, in ~\cite{paper_87}, the authors propose a parsimonious architecture by selecting how many microservice instances the overall system must have regarding resource provision. The authors emphasize that adopting CPU utilization as the key metric to determine the instance quantity is not sharp enough. The architecture might employ other metrics, like microservice metrics (request time, wait time, requests served per second) or alternative ones from the RED Method \cite{redmethod2018site}. 

In addition, resource management is strongly related to the type and amount of requests the system receives. Consequently, some studies profile the requests to achieve better resource utilization. They profile each request's network, energy, CPU/GPU, RAM, and QoS requirements to select the edge nodes that better fulfill these profiles \cite{paper_94,paper_91}. Also, these resources have to be monitored for status and health observation. In ~\cite{paper_30,paper_1}, an agent controls and monitors resource usage, node status, and other architecture parameters to provide the orchestrator with valuable information to manage and offload the microservices. Further, the authors in \cite{paper_1,paper_22,paper_17}  employ a time-series database (e.g., InfluxDB) to store architecture health information and enable periodical assessment. 

Every microservice could be designed and implemented using different technologies/programming languages in an architecture. Consequently, a microservice must be capable of working with a plethora of different technologies, a concept called \texttt{polyglot microservice} \cite{buildingmicroservices_samnewman_book}.
Thus, several database systems can be used in a microservice architecture, like Cassandra \cite{paper_1}, InfluxDB \cite{paper_34,paper_29}, MySQL \cite{paper_6,paper_20}, MongoDB \cite{paper_10,paper_22}. Moreover, the microservices are developed using different programming languages and frameworks such as, Python \cite{paper_6,paper_20,paper_23,paper_25,paper_34,paper_48,paper_52,paper_61,paper_64,paper_81,paper_77,paper_87}, Java OSGi \cite{paper_4,paper_18}, SpringBoot \cite{paper_7,paper_25,paper_47}, Java \cite{paper_8,paper_22,paper_23,paper_43,paper_84,paper_82}, Golang \cite{paper_10,paper_25,paper_54}, Node.JS \cite{paper_18,paper_36}, C++ \cite{paper_20,paper_23,paper_64}, R \cite{paper_26}, and even pure C \cite{paper_33}.

Since it is required to work with different technologies in the same architecture, a common communication interface among the microservices is necessary. Hence, many architectures use the \texttt{JSON} pattern to exchange messages among the microservices while performing orchestration, offloading, placement, and synchronization ~\cite{paper_10,paper_17}. \texttt{JSON-LD (JSON for Linking Data)}, an extension of the original \texttt{JSON} to provide better context information, is applied in \cite{paper_4}. Moreover, many message protocols are employed in architectures using \texttt{JSON}, like MQTT \cite{paper_8,paper_28,paper_34}, CoAP \cite{paper_25}, and LoRa \cite{paper_34}. Finally, microservice architectures explored several enterprise tools, like the Apache ecosystem: Kafka, Camel, and Storm~\cite{paper_7}.

Different features from other well-established technologies (e.g., Multi-agent Systems, Holonic Systems\footnote{The holonic systems comprise a system composed of \textit{holons}. The holon is an independent entity that could work alone (atomic) or in a group forming a \textit{holarchy}. The holarchy is seen from the outside as a singular unity, however, inside it is a group of holons. Holonic systems have been applied in different scenarios, such as autonomous systems and manufacturing systems.})) have been employed in the microservice design to improve and enhance its functionalities. Microservices can be reactive based on agent characteristics \cite{paper_40} to achieve high availability for IoT applications by interacting with a reactive broker. Cognitive agents \cite{paper_50} can support microservice orchestration, helping deal with the dynamicity of this process. In \cite{paper_59}, agents and microelements\footnote{This concept comes from the Osmotic Computing, where the microelements are an encapsulation of microservice and data} support orchestration. Finally, agents could provide a model to deploy the microservices on the Edge by following a BDI model. In ~\cite{paper_75}, the authors employ the agent model to provide continuous integration when microservices get updated or deleted.

The \texttt{cyber-microservice} concept was proposed in \cite{paper_18} and consisted of a microservice encapsulating sensors and actuators in a single entity. Nanoservices are miniature microservices focusing on the IoT scenario \cite{paper_20,paper_23,paper_53}. The nanoservice is a lower-size microservice capable of running on a resource-constrained embedded platform (Raspberry Pi 2/ESP 8266), aiming at mobility among the nodes. Nikouei et al. \cite{paper_21} proposed a microservice with a variable granularity, where its size changes according to architecture demand, scenario, or even the platform. While the microservice size changes, they can combine functions, which will work together and be seen as a single service. 

Developing distributed systems like microservices requires design patterns to produce robust, standardized, and scalable systems~\cite{microservice_designpatterns}. Table \ref{tab:slr_patterns} shows the design patterns explored by the selected studies.\footnote{Although we identified other design patterns in the selected studies, we listed on the Table \ref{tab:slr_patterns} only the studies that explicitly described the pattern usage}.

\begin{table}[h]
\centering
\small
\begin{tabular}{@{}l|l@{}}
\toprule
\textbf{Design Pattern} & \textbf{Reference} \\ \midrule
Query                   & \cite{paper_7,paper_17}               \\ \midrule
Sidecar                 & \cite{paper_8,paper_30}               \\ \midrule
Backend for Frontend    & \cite{paper_30}               \\ \midrule
Service Mesh            & \cite{paper_30,paper_33}                  \\ \midrule
Health Check            & \cite{paper_30,paper_40}               \\ \midrule
Service Discovery       & \cite{paper_30,paper_96,paper_40,paper_45}               \\ \midrule
Circuit Breaker & \cite{paper_40} \\ \midrule
Downshifting & \cite{paper_45} \\ \midrule
Monitoring Connector & \cite{paper_45} \\ \midrule
Edge-to-Cloud Balancing & \cite{paper_45} \\ \midrule
Test Probe & \cite{paper_45} \\
\bottomrule
\end{tabular}
\caption{Design Patterns}
\label{tab:slr_patterns}
\end{table}

In this review, we consider API Gateway as an architectural feature, as listed in Table ~\ref{tab:slr_features}, but it is also a design pattern \cite{microservice_designpatterns}. The \texttt{Query} pattern is based on the API Gateway and consists in querying an incoming request from the user or system and splitting it into queries to the respective microservices. This pattern is also called the \texttt{API Composition pattern} and could be carried by the API Gateway microservice in some architectures \cite{paper_17}. Another pattern based on the API Gateway is the \texttt{Backend for Frontend}. It employs different API Gateways along the architecture to handle user requests based on their interests, groups, and purposes \cite{paper_30}.

Some patterns aim to handle different microservices implementations in the same architecture, such as \texttt{Service Mesh} and \texttt{Sidecar Pattern}. As explained in the Microservice-Design Patterns book ~\cite{microservice_designpatterns}: "The \texttt{Service Mesh} is a networking infrastructure that mediates the communication between a service and other services and external applications. All network traffic in and out of service goes through the service mesh. It implements various concerns, including circuit breakers, distributed tracing, service discovery, load balancing, and rule-based traffic routing". Together with the \texttt{Service Mesh}, the \texttt{Sidecar pattern} places a sidecar object running aside the microservices to provide the concerns from the \texttt{Service Mesh} pattern. The \texttt{Service discovery} pattern works in two modes: client-side discovery and server-side discovery. In client-side discovery, the client is in charge of discovering the microservice running instances and invoking them. On the other hand, in server-side discovery, the client requests a router (e.g., an API Gateway) for a specific function, and then the router invokes the microservices \cite{microservice_designpatterns,paper_96}. Finally, the \texttt{Health Check} pattern is an endpoint (/health) that each microservice provides to inform its health status. 

In \cite{paper_45}, the authors describe patterns to cover the four phases of industrial software's life cycle to Edge Computing: \textit{deployment, monitoring, adaptation}, and \textit{testing}. For \textit{deployment}, the \texttt{Downshifting} is proposed to move from Cloud to the Edge, a microservice that needs a lower response time. The \texttt{Monitoring Connector} proposes \textit{monitoring} the microservice's component interaction to avoid malicious or unwanted behaviors. In the \textit{adaptation} phase, \texttt{Edge-to-Cloud Balancing} provides a load balance between Edge and Cloud to handle the scalability between the tiers. Lastly, in the \textit{testing} phase, the \texttt{Test-Probe} pattern proposes reliable tests, accessing internal data of microservices in order to maintain them in production.

\begin{table}[h!]
\centering
\small
\begin{tabular}{@{}ll@{}}
\toprule
\textbf{Platform}                      & \textbf{Reference}                                    \\ \midrule
Container                              & \begin{tabular}[c]{@{}l@{}}
\cite{paper_58,paper_41,paper_98,paper_26,paper_10,paper_78,paper_27,paper_38,paper_95,paper_102,paper_87,paper_22,paper_8,paper_16,paper_65,paper_80,paper_61,paper_6,paper_1,paper_7,paper_11,paper_15,paper_20,paper_23,paper_48,paper_51,paper_53,paper_54,paper_60,paper_70,paper_71,paper_72,paper_73,paper_75,paper_79,paper_82}\\
\cite{paper_85,paper_88,paper_89,paper_90,paper_93,paper_97,paper_99,paper_108,paper_17,paper_28,paper_40,paper_42,paper_47,paper_77,paper_5,paper_18,paper_33,paper_35,paper_43,paper_59,paper_62,paper_32,paper_36,paper_44,paper_52,paper_64,paper_74,paper_81,paper_109,paper_0,paper_29,paper_107,paper_21,paper_67,paper_76}

\end{tabular} \\ \midrule
Independent&
\begin{tabular}[c]{@{}l@{}}
\cite{paper_24,paper_46,paper_94,paper_104,paper_84,paper_31,paper_39,paper_2,paper_9,paper_12,paper_14,paper_30,paper_55,paper_68,paper_83,paper_86,paper_91,paper_92,paper_96,paper_105,paper_106}\\
\cite{paper_49,paper_50,paper_57,paper_63,paper_69,paper_100,paper_101,paper_56,paper_66,paper_110,paper_19,paper_34,paper_45,paper_3,paper_25,paper_37}
\end{tabular}
\\ \midrule
Virtual Machine                        &             \cite{paper_98,paper_13,paper_53,paper_4}                                            \\ \bottomrule
\end{tabular}
\caption{Platform}
\label{tab:slr_platform}
\end{table}

Microservices are capable of working on different platforms. Table \ref{tab:slr_platform} lists the three platforms used in the selected studies: containers, independent\footnote{It means that the study does not specify which platform hosts the microservices.}, and virtual machines. Most architectures run Docker containers due to their lightweight, isolation, and flexibility \cite{paper_1,paper_26,paper_58}. In \cite{paper_11}, the authors discuss how many microservices instances should run in each container. The authors conclude that the best suitable option is running n-microservices instances per container instead of a single instance, aiming at system scalability. On the other hand, two patterns base their claim: the one-microservice-per-host and multiple-service-per-host patterns \cite{microservice_designpatterns}. 

Regarding the nodes, different devices (embedded/virtual) run as Edge Nodes (servers and/or hosts) along the microservice architectures. These devices are an excellent match to the microservices' characteristics (e.g., lightweight), aiming to accomplish architecture tasks (e.g., orchestration and offload). In Table \ref{tab:slr_devices}, we list the devices employed by the selected studies to run as edge nodes.

\begin{table}[h!]
\centering
\small
\begin{tabular}{@{}ll@{}}
\toprule
\textbf{Device}         & \textbf{Studies}                                       \\ \midrule
Raspberry Pi 1          & \cite{paper_82} \\ \midrule
Raspberry Pi 2          & \begin{tabular}[c]{@{}l@{}}\cite{paper_20,paper_23,paper_42}\end{tabular}          \\ \midrule
Raspberry Pi 3          & 
\begin{tabular}[c]{@{}l@{}}
\cite{paper_58,paper_10,paper_27,paper_87,paper_22,paper_8,paper_94,paper_1,paper_15,paper_20,paper_23,paper_54,paper_70,paper_79,paper_83,paper_99,paper_40,paper_77,paper_34,paper_52,paper_109,paper_25}\\
\cite{paper_29,paper_21}
\end{tabular} \\

\midrule
Raspberry Pi 4          & \begin{tabular}[c]{@{}l@{}}
\cite{paper_98,paper_103,paper_94,paper_11,paper_70,paper_93,paper_101,paper_25,paper_107}\end{tabular} \\ \midrule
Beaglebone Black (BBB)  & \cite{paper_64} \\ \midrule
Custom Hardware         & \begin{tabular}[c]{@{}l@{}}\cite{paper_16,paper_7,paper_14,paper_53,paper_83}\end{tabular} \\ \midrule
Intel NUC               & \cite{paper_79,paper_64} \\ \midrule
Lenovo ThinkCentre 920x & \begin{tabular}[c]{@{}l@{}}\cite{paper_6,paper_83}\end{tabular} \\ \midrule
Nvidia Jetson TX2       & \begin{tabular}[c]{@{}l@{}}\cite{paper_103,paper_9,paper_17,paper_107}\end{tabular} \\ \midrule
Nvidia Jetson Xavier & \cite{paper_101} \\ \midrule
Google Coral Dev AI Board & \cite{paper_110} \\ \midrule
Simulated (Amazon EC2)  & {\begin{tabular}[c]{@{}l@{}}
\cite{paper_12}\end{tabular}} \\ \midrule
ATMega 2560             &       \cite{paper_20,paper_23}                                                 \\ \midrule
ThinkerBoard            &       \cite{paper_22}                                                 \\ \midrule
Smartphone             &       \cite{paper_37}                                                 \\ \bottomrule
\end{tabular}
\caption{Selected Studies - Edge Node - Devices}
\label{tab:slr_devices}
\end{table}

\subsection{RQ1.1 - Based on the microservice architectural approaches and features, are there advantages of using microservices in the Edge?}

The selected studies highlight several advantages, primarily because Edge Computing is a decentralized paradigm next to the users' devices. Hence, applying a microservice architecture leads to a suitable merge of these technologies. However, it leads also to some research challenges and issues that have to be handled by the microservice architecture developers. 

\begin{itemize}
    \item \textit{Lower response time:} Developing and deploying in the Edge provides a lower response time for the microservices to handle the requests than in the Cloud. It impacts IoT applications committed to a very low response time (e.g., smart car) or a minimum QoS level to their users (e.g., real-time video surveillance). For instance, an IoT application that controls a servo motor in a factory could not handle a variable or high response time \cite{paper_7,paper_12}.

    \item \textit{Distributed Computing:} Edge Computing provides several edge nodes, and each microservice carries a single business capability. Thus, having multiple Edge nodes to assign microservices is an advantage to a microservice architecture. Moreover, having the distributed nodes along the Edge enables the architectures to split the heavy work among the edge nodes by employing many microservice instances and controlling them in a superior tier (Fog or Cloud). Video streaming architectures \cite{paper_9} explored this perspective, where the render process occurs in the edge nodes, the orchestration in a fog node, and the storage in the Cloud. As a distributed paradigm, node control and discovery are important features that these microservice architectures must provide. In \cite{paper_32}, the authors use the DNS system to discover available nodes to host the microservices based on smart contracts. Additionally, Edge's distributed computing provides better microservice chaining due to the node proximity \cite{paper_6} in a choreographic architecture.

    \item \textit{Lightweight:} The microservice lightweight characteristic enables the architectures to provide scalability mechanisms in the Edge. First, offloading the microservices among the edge nodes (horizontal) and the tiers (vertical). This microservice "flow" among tiers has been proposed by the concepts of Capillarity Computing \cite{paper_1} and Osmotic Computing \cite{paper_13,paper_20} as well. Second, this feature also improves the orchestration and choreography due to easier deployment, placement, invocation, and coordination of lightweight microservices instead of monolith applications \cite{paper_6,paper_15}.
  
    \item \textit{Resource management:} Distributing and offloading the microservices along the Edge instead of sending them directly to the Cloud provides better network and resource utilization \cite{paper_7,paper_9}.
    
    \item \textit{Decomposition:} Splitting the microservices along an n-tier architecture improves the desired trade-off between flexibility and performance by having distinct tiers to handle different demands \cite{paper_8}.
\end{itemize} 

\subsection{RQ1.2 - From the architectural perspective, which microservice characteristics take advantage of Edge Computing usage?}

Some microservice characteristics are enhanced when employed in the Edge Computing scenario. Below we describe some of these features:

\begin{itemize}
    \item \textit{Single Business Capability:} Whereas each microservice carries a single business capability, it takes advantage of the Edge distributed scenario by deploying the microservices along the edge nodes.
    
    \item \textit{Social aspect:} While working in the Edge environment, the microservices share a common message interface to coordinate, cooperate, and synchronize their peers along the edge nodes. Besides the message interface, a polyglot capability is desired for a microservice architecture to handle different implementations and technologies \cite{paper_25}. 
    
    \item \textit{Throughput and performance:} In \cite{paper_20}, the authors highlight the throughput feature of a microservice architecture when deployed in an Edge Computing scenario mainly due to the proximity to the users, thus avoiding a network bottleneck when compared to a Cloud-only scenario. Besides distributed computing, microservices gain performance when managed, composed, and deployed in the Edge due to fewer network hops so the user can access the services. 
    
    \item \textit{Running in resource restricted platforms and Isolation:} Whereas the microservice works as an isolated entity in a platform (virtual machine, container) being able to handle a business capability, it is possible to run it on a restricted platform (e.g., an embedded platform) as seen in Table 14 from the Research Question 1. 
    
    \item \textit{Service availability:} Whereas the microservices are available in the Edge tier, it enhances the overall system availability by eliminating eventual access problems to the core networks \cite{paper_20}.
\end{itemize}

\subsection{RQ2 - What are the fundamental microservices characteristics to properly work in Edge Computing?}

The selected studies pointed out some fundamental microservice characteristics to properly work in Edge Computing. In this context, the term \textit{properly} means guaranteeing the Quality of Service (QoS) to the users. Microservice architectures apply different QoS metrics. Most of these metrics are based on the network infrastructure, like the number of hops, bandwidth, packet loss, throughput, delay, and jitter \cite{paper_1}, or on microservice metrics, like the number of requests served per second, the number of failed requests per second, and how long it takes to handle a request\footnote{The RED Method employs these metrics in order to measure how a microservice architecture is working \cite{redmethod2018site}}.

\begin{itemize}
    \item \textit{Lightweight:} A microservice is inherently micro. However, being lightweight is even more crucial \cite{paper_13} due to the Edge characteristics. 
     
    \item \textit{Light Communication:} A microservice should interact through a light messaging mechanism. The messages are synchronous or asynchronous. For instance, most microservice architectures employ an asynchronous message style using JSON or JSON-LD patterns~\cite{paper_10,paper_5,paper_17,paper_28}. Nevertheless, some architectures also employ synchronous messages like RPC (Remote Procedure Call) \cite{paper_14,paper_24}.
    
    \item \textit{Loose coupling and scalability:} Scalability performs a fundamental role in assuring the service's continuous delivery to the users as the demand increases, and may work in two ways: \textit{up} and \textit{down} \cite{paper_30,paper_15}. The scale-up mode aims to increase the number of microservice instances and, consequently, the resources to meet the increasing demand. On the other hand, whenever the demand goes down, the architecture turns off unused microservice instances to decrease resource utilization. Therefore, scalability is directly related to the microservice operational cost. Microservices are loosely-coupled to be scalable. In other words, an orchestrator can assign them to different edge nodes. 
    
    \item \textit{Trust:} Several microservices spread along the edge nodes providing different demands and purposes. Therefore, the microservice architecture must guarantee that only trustful microservices are allowed. Blockchain \cite{paper_1,paper_15,paper_30}, encryption \cite{paper_36}, Cloudlet proxy \cite{paper_16}, and public/private key \cite{paper_24} are among the mechanisms to support that.
    
    \item \textit{Interoperability:} Several programming languages provide support for microservices development. Microservice architectures rely on different programming languages and technologies concurrently running at the same node. Therefore, architectures must work towards integration among the microservices by providing standard interfaces and interoperability features \cite{paper_25}.
    
\end{itemize}

\subsection{RQ3 - What are the techniques employed in the microservice composition process? In which tiers do they work?}
\label{subsec:rq3}

Several selected studies employ the microservice orchestration by using a container orchestration tool, such as Docker Swarm \cite{paper_5} or Kubernetes \cite{paper_9,paper_13,paper_35}. However, as pointed out by the authors in \cite{paper_24} and \cite{paper_38}, the current container orchestration tools do not serve as expected to face the dynamicity of the Edge Computing scenario.

\begin{quote}
    \textit{``Orchestrators for containerized MSA [Microservice Architecture] applications exist. However, these orchestrators do not scale to the number of nodes that will be required in this new paradigm, nor are they designed to work in the EC [Edge Computing] scenario.''}  - \cite{paper_24} \\ and furthermore in \cite{paper_38}: \textit{``Current container orchestrators, such as Kubernetes, are centralized solutions meant for managing local clusters in public and private cloud settings. They have a limited ability to scale, and assume high-performance network connectivity between the nodes, i.e., low latency, high bandwidth and low packet loss.''} 
\end{quote}

So, from now on, we focus on orchestration mechanisms specifically developed for microservices in the Edge Computing scenario. Let's analyze the main types of microservice composition:

\begin{itemize}
    
    \item \textit{n-tier microservice orchestration:} It is required to define which tier hosts the orchestrator service. The orchestrator could be in the same tier as the microservices (Edge server) \cite{paper_24,paper_26,paper_32}. Alternatively, it can work in a superior tier, Fog or Cloud~\cite{paper_9}, or even in an intermediate layer between the tiers provided by a sidecar object~\cite{paper_8}. Some studies are based on the Cloud-Edge Continuum concept. In this approach, the orchestrator must be able to handle the microservices along the Continuum ~\cite{paper_98,paper_102}. The authors in \cite{paper_102} and \cite{paper_95} introduce some of the main challenges that an orchestrator in the Continuum must be aware of: i) communication delay through a public network. Hence, it requires a map between Edge and Cloud to define routes among nodes properly; ii) Establish a well-known resource policy to avoid node bottlenecks or even waste. For that, the architecture must map every hardware structure (CPU, GPU, RAM, and disk). Some monitoring tools could produce this hardware map, for instance, Grafana \cite{grafanasite}; iii) Microservice seamless execution in the Continuum. The orchestrator must guarantee a seamless execution to the users, satisfying the QoS requirements, even when microservice instances change tiers. The orchestrator could play a role deeper than simply managing the microservice instances running (or being prepared to run) in the Edge/Fog/Cloud nodes. Some studies propose an orchestrator in charge of resource scheduling employing machine learning techniques (e.g., differential evolution) to predict resource availability \cite{paper_97}.

    \item \textit{Orchestration and API Gateway:} The Orchestrator could work with the API Gateway to achieve stronger cooperation in the service composition process. Moreover, in \cite{paper_20}, the authors assure that API Gateway and Orchestrator could work as a single entity. 
    
    \item \textit{Decentralized Orchestration:} It consists of microservices that do not follow a centralized and hierarchical orchestrator. The microservices can work in a full-flat scenario, where every single microservice may orchestrate its peers or organize temporary coalitions to answer a given request \cite{paper_38}. For instance, the authors in \cite{paper_24} propose a decentralized orchestration utilizing different roles: leader, host, controller, and entry. When a request comes to the system, a group of at least four microservices is designated to process the request by adopting those roles. The entry microservice answers the request. After, one or n-controllers are assigned to ensure that a leader will rule the request. Thus, the controllers elect a leader. Finally, the leader chooses a host to receive the application requested, and then the host regularly sends to the leader its health status. When the request finishes, the microservices leave their roles to fulfill subsequent requests. 
    
    \item \textit{Meta-Orchestration:} An orchestrator may become too complex to maintain when dealing with heterogeneous infrastructures and environments. Meta-orchestration proposes to split the orchestration task into two layers: meta and low-level. The low-level layer comprises n-orchestrators, each handling a specific environment and infrastructure. Additionally, the meta-orchestrator coordinates the low-level ones \cite{paper_101}.
    
    \item \textit{Choreography:} Although choreography has been cited as a great mechanism to employ in microservice architectures ~\cite{microservicespatternsBook,buildingmicroservices_samnewman_book,singhal2019orchVSchor,eucapaliptoolvalderas2020} aiming at decentralized microservice composition, it appeared only in two selected studies. In ~\cite{paper_2}, the authors analyze how a choreography mechanism impacts system requests. Also, they use a formal method to verify the choreography called \texttt{chor2norms}. The second study employs choreography~\cite{paper_10} to a WIA-PA (Wireless Networks for Industrial Automation- Process Automation) gateway specific to the Edge. The microservices work as a distributed architecture communicating via a message bus. 
    
    \item \textit{Hybrid Microservice Composition:} Some authors have been proposing to employ a hybrid microservice composition, in other words, to have orchestration and choreography working simultaneously \cite{eucapaliptoolvalderas2020}. The reason to work towards a hybrid approach is the Edge Computing scenario dynamicity. The Edge constantly changes: new edge nodes, variable demand, and distinct QoS requirements. This hybrid microservice composition could be supported and invoked by switching the microservice architecture topology. Whenever the topology switches, the composition switches as well \cite{paper_85, paper_87}.
    
\end{itemize}

\subsection{RQ4 - How does the microservice offloading process work?}

There are three main aspects to offloading: when to offload, where to offload, and how to implement the offloading. 


Regarding where to offload, the offloading process is divided into two groups: vertical and horizontal. Vertical refers to offloading a microservice to a superior tier, like Edge to Fog or Edge to Cloud. On the other hand, horizontal offloading occurs in the current microservice tier. 

Vertical offloading is usually motivated by seeking a node with higher computational resources (hardware and/or network bandwidth) in a superior tier \cite{paper_1,paper_9}. For example, some microservice architectures that employ neural networks through Deep Learning offload the training model to run in a fog or cloud node seeking a higher processing power, and the inference model stays on the edge node~\cite{paper_55}. Vertical offloading usually works in explicit mode, where the microservice moves to another host. However, the offloading process may also occur implicitly, where Edge, Fog, and Cloud are seen as the same tier, as in Continuum Computing~\cite{milojicic2020edgecontinuum}, pipeline \cite{paper_33,paper_53}, Capillary Computing \cite{paper_1}, or Osmotic Computing \cite{paper_13}.

The horizontal offloading is mainly motivated by the dynamicity of the Edge (unavailable nodes, network instability, geographical constraints, user mobility, and so on). Although, it may also be motivated by seeking a higher capacity node if the Edge computing scenario offers it. If there are distinct players (server, gateway, nodes(hosts)) in the Edge, the horizontal offloading can work among the nodes or into the server or gateway \cite{paper_25}.

Furthermore, both offloading mechanisms could simultaneously exist in the same architecture (Edge to Edge and Edge to Fog/Cloud model) \cite{paper_10,paper_87,paper_109}, giving the option to offload the microservice to Edge peers or a higher tier. One example is Mobile Edge Computing (MEC) \cite{paper_85}, where mobile devices could work as host nodes and run microservices offloaded from their peers seeking lower latencies. But depending on how latency-sensitive a microservice is, it is necessary to determine in run-time where to offload it: Edge or Cloud. The authors argue that it is preferable to offload on the Edge to reduce latency, to the detriment of performance. In \cite{paper_87}, the authors propose different architecture topologies (hierarchical, master-slave, mesh, client-server). Hence, the offloading choice depends on the topology. 

The offloading decision could be carried by an orchestrator or the microservice itself. Different techniques to select the proper tier and node to host the microservice have been explored. In \cite{paper_109}, the authors employ a Markov Decision Process (MDP) combined with a Support Vector Regression (SVR) to select the offloading target node. Also, in \cite{paper_73}, the authors formulate a Fuzzy Logic based on hardware parameters to support the offload decision. Whereas this decision process consumes computational resources, it can be performed in a superior tier (e.g., Fog) \cite{paper_52}. Therefore, each microservice architecture has to deal with this dilemma: Where does the microservice go? The appropriate answer is: it depends on the application QoS requirements. 

The offloading process consists mainly of sending the microservice to another tier/host aiming, for example, higher computational power or lower delay network to achieve the desired QoS. Different techniques could be applied in microservice offloading. For instance, instead of temporally offloading a whole microservice to another host, only part of the microservice might be offloaded, like a thread or a function \cite{paper_6}. 

Further, there is a need to ensure that the new host will provide the required performance. In other words, the architecture must face the challenge of guaranteeing performance among the nodes. For example, in \cite{paper_68}, FPGAs are employed in a Cloud cluster to receive the offloaded services to accelerate their execution. Nevertheless, some authors proposed a different offloading mechanism: microservice store \cite{paper_6}. The store could be in an edge node and keeps the microservice binary. Whereas there is a request to an unavailable microservice (e.g., offline or unreachable), the system provides a new instance and offloads it to an edge node. The authors argue that the store technique overcomes the traditional offloading mode by saving energy and reducing latency. 

\subsection{RQ5 - What challenges/problems are identified in the research literature relating to microservices and Edge Computing?}


The selected studies relate different challenges and problems when a microservice architecture runs in Edge Computing. Below, we describe some of the most pointed out by the authors that should be cared for when we build solutions using microservices on Edge Computing. 

\begin{itemize}
    \item \textit{Container and microservice orchestration/choreography:} Choosing which approach (orchestration or choreography) to adopt is a challenge that every microservice architecture shall face to manage several microservices running simultaneously along with the nodes. Both approaches have pros and cons. Orchestration brings a centralized entity that coordinates the whole architecture, leading to less complexity to deploy and analyze. On the other hand, choreography brings flexibility by working decentralized, while it carries a higher complexity. Therefore, combining these service compositions (a hybrid version) is a clever solution to design and deploy in a microservice architecture \cite{singhal2019orchVSchor,eucapaliptoolvalderas2020}.
    
    Regarding the comparison of microservice and container orchestration, commercial container orchestration does not focus on Edge Computing dynamicity and distribution \cite{paper_24,paper_38}. Thus, there is a need to build robust, scalable, and flexible solutions to handle the microservice orchestration and choreography in Edge Computing. 
    
    \item \textit{Deployment and maintenance:} Updating or replacing a running microservice is a challenge that has to be dealt with, whereas the whole architecture keeps up. For example, the sidecar pattern proposed in [41] provides the so-called sidecar object to maintain the microservice in production time. Another solution to this issue is employing microservice instance replicas throughout the nodes \cite{paper_35,paper_38}. Finally, a microservice store, as explored in \cite{paper_6}, could work to provide fresh replicas when facing a maintenance update. 
    
    \item \textit{Placement:} When the architecture starts, placing the microservices in an n-tier architecture provides higher flexibility due to the node's availability. However, it leads to the microservice placement challenge. Different approaches are proposed to provide placement mechanisms, from relying upon the container orchestration tools \cite{paper_5} to an automatic placement configuration based on the system context and status \cite{paper_12}.

    \item \textit{Scalability}: Whereas the architecture goes over time, it could be required to scale node and microservice availability vertically or/and horizontally. Regarding microservices, the scalability up and down is an important feature that an MSA should have when running in an Edge Computing scenario \cite{paper_88,paper_90}. Further, when the architecture faces constrained resource availability, energy saving is an enabler to keep the whole system up. Hence, the scale-down is essential in those scenarios to save energy and avoid resource loss. 
    
    \item \textit{Security and privacy:} Some of the selected studies' most significant challenges were about security and privacy in the Edge. Unlike the Cloud, where the data and microservices are located in far-away controlled servers, Edge Computing offers a distributed scenario closer to the users. It leads to a privacy and security issue that has to be faced. Firstly, a trust mechanism through blockchain among the microservices is used in \cite{paper_0,paper_13,paper_15} by providing a trust value to assess how trusted is a microservice and an edge node. This trusteeship among the microservices might be managed by employing smart contracts \cite{paper_97}. Secondly, facing the distributed scenario, an authentication mechanism is required to ensure the microservice instances' integrity \cite{paper_85}. Finally, an architecture might assure the security of microservice data and transactions by protecting data utilizing AES/TKIP encryption algorithms. The authors in \cite{paper_93} propose a secure gateway, a secure version of an API-Gateway, which is in charge of encrypting all the microservice interactions. 
    
 Deploying Edge Computing in an old-fashioned platform also implies handling security issues by exposing these platforms to a network for connecting directly to the Cloud, for instance. Employing an edge node (cloudlet) between the platform and Cloud may work as a firewall and proxy mechanism, while this node also works as an offloading station \cite{paper_16}. In addition to the firewall, encrypting the microservice data is a challenge handled in \cite{paper_36}, where the edge nodes encrypt using an ECDSA algorithm to assure security. However, it leads to another challenge: performance \cite{paper_36}.
    
\end{itemize}

\section{Conclusion}
\label{sec:conclusion}

Putting a Microservice Architecture facing Edge Computing brings an opportunity for this architecture to work in a distributed, low-latency, heterogeneous, and complex scenario. After 2014, when the Microservice term gained the spotlight, a considerable amount of research has been published pointing the Microservices as the suitable architecture to work in Edge Computing.

As discussed in this work, architecture approaches and features, composition, and offloading are the main aspects of microservice architectures in the Edge. To better understand the current state of those aspects,  we conducted a Systematic Literature Review (SLR). The goal was to identify relevant research relating to Edge and Microservices. We focused on studies that present the following: i) microservice architecture approaches and features applied and designed to work in Edge Computing; ii) microservice orchestration/choreography techniques and tools focusing the Edge; iii) Offloading methods applied to Microservices in the Edge. The SLR selected, from 1818 papers, 111 studies to be carefully analyzed and classified according to these three aspects. We also built a taxonomy (see Fig \ref{fig:slr_taxonomy} in Section \ref{subsec:rq4}) where the selected studies were classified. 

While conducting the SLR analysis, some topics gained our attention as trends and research gaps. Following, we describe them in detail. 

\textbf{Orchestrating microservices.} We highlight this subject both as a research gap and a trend. As a research trend, 21 studies employ some orchestration in their projects. As a gap, Section \ref{subsec:rq3} discusses the demand for specific microservice orchestrators designed for Edge Computing instead of using container orchestrators.


\textbf{Choreography.} Microservice architectures tend to employ a decentralized way to coordinate microservices, as pointed out by Martin Fowler’s book~\cite{buildingmicroservices_samnewman_book,microservicespatternsBook}. Therefore, choreography could be a popular microservice composition method in the studies. However, as we have seen in the SLR, this is not the reality. We believe it occurred for several reasons, such as i) The choreography concept is not as well-known as orchestration; ii) Some authors handle the decentralized nature by designing a decentralized orchestration~\cite{paper_38,paper_24}; iii) Due to the  orchestration gap identified above since container orchestrators cannot support choreography;

\textbf{Microservice design patterns.} The selected studies present many design patterns usages. However, only a few studies describe the design pattern implementation. In many other cases, design patterns become essential features (e.g., API Gateway, Service Discovery). Therefore, we see the design patterns as a trend in microservice architectures.


\textbf{Auxiliary systems.} The study demonstrated a plethora of microservice architectures focused on Edge computing. Several architectures have followed a trend by employing auxiliary/support systems to enhance and empower the whole architecture. We understand this trend as threefold: i) The security/privacy by employing blockchain; ii) The overall performance and architecture health monitoring, by node and microservices, through time-series DB and monitoring tools (e.g., Grafana); iii) Multiagent Systems to enhance the IoT layer managing and microservice interaction.

\textbf{Microservice architectures approaches and features.} The approaches that stood out in the studies were the 1-tier (Edge) and 2-tier (Edge + Cloud) architectures. Moreover, some authors employ an extra tier presenting a learning feature applied in orchestration and offloading. Regarding architecture features, the API Gateway is present in 18 of the 111 selected studies. It provides an entry point to the system, load balance, and even nests an orchestrator to the microservices. Therefore, we may classify the API Gateway as a trend in the microservice architectures in Edge Computing. 

\textbf{Edge Devices.} The Raspberry Pi family devices (RP3 and RP4) are the most utilized Edge Devices in the microservice architectures. The main reasons are adequate performance and low cost. Both systems deliver enough processing power to run as Edge devices, even more when employed in an n-tier architecture. Indeed, there is a remarkable difference between RP3 and RP4 in computing power, as demonstrated in several benchmarks available on the Internet\cite{rp3rp4perfomanceSite,rp3rp4perfomanceSite2}. The cost of both devices on the release date was around 35 U.S. dollars, which is an attractive price to work in a distributed scenario like Edge Computing. Other devices have also been applied, such as the NVidia Jetson family devices (TX2 and Xavier). These devices surpass the Raspberry Pis’ performance. However, they cost next to 399 U.S. dollars\footnote{This is a released price from an NVIDIA News/Blog website (2020) \cite{nvidiaJetsonXavierrelease}}.

\textbf{Offloading methods.} We recognize offloading as a research trend due to 17 studies employing offloading in their architectures to seek a higher performance in a superior tier or a lower response time to accomplish a satisfactory QoS.

Summarizing, we conclude the SLR containing 111 selected studies indicating several research trends and gaps in the current state of the art. 
This SLR will help academic researchers and industry professionals to get an overview of Microservices on Edge Computing in terms of architecture approaches and features, microservice composition, and offloading.

\section*{Acknowledgments}
\label{sec:ack}
This research was supported by Coordena\c{c}\~{a}o de Aperfei\c{c}oamento de Pessoal de N\'{i}vel Superior - Brasil (CAPES - Finance Code 001) and QuintoAndar. The authors are grateful for the financial support.

\section{Appendix A. List of papers and quality scores}
\label{sec:appendixA}

\setlength{\tabcolsep}{2.5pt}

\begin{table}[h]
\small
\begin{tabular}{@{}llllllllllllllllll@{}}
\cmidrule(lr){3-15}
                                  &             & \multicolumn{13}{c}{\textbf{Quality Questions}}                                                                                                                            &                                                                &                  &                    \\ \midrule
\multicolumn{1}{c}{\textbf{Reference}} & \multicolumn{1}{l}{\textbf{RT}} & \textbf{1}           & \textbf{2}           & \textbf{3}           & \textbf{4}           & \textbf{5}           & \textbf{6}           & \textbf{7}           & \textbf{8}           & \textbf{9}           & \textbf{10}          & \textbf{11}          & \textbf{12}          & \textbf{13}          & \textbf{\begin{tabular}[c]{@{}c@{}}Total\\ Score\end{tabular}} & \textbf{Qual.}     & \textbf{Cit.} \\ \midrule
(Zheng et. al, 2021) \cite{paper_0}                           & SP          & 0          &            & 1          & 1          & 0.5        & 0.5        & 0.5        &            & 1          & 0.5         &             &             & 1           & 6                                                              & 66.7             & 13                 \\
(Taherizadeh et. al, 2018) \cite{paper_1}                            & SP          & 1          &            & 1          & 1          & 0.5        & 1          & 0.5        &            & 1          & 0.5         &             &             & 1           & 7.5                                                            & 83.3             & 95                 \\
(Fei et. al, 2020) \cite{paper_2}                            & ER          & 0          & 1          & 1          & 1          & 1          & 1          & 1          &            & 1          & 0.5         &             &             & 0.5         & 7.5                                                            & 75               & 22                 \\
(Bixio et. al, 2020) \cite{paper_3}                            & SP          & 0.5        &            & 1          & 1          & 1          & 1          & 1          &            & 1          & 0.5         &             &             & 0           & 4                                                              & 44.4             & 5                  \\
(Kaneko et. al, 2020) \cite{paper_4}                            & SP          & 0.5        &            & 1          & 1          & 1          & 1          & 1          &            & 0.5        & 0.5         &             &             & 0.5         & 7                                                              & 77.8             & 2                  \\
(Nikolakis et. al, 2020) \cite{paper_5}                            & SP          & 0.5        &            & 1          & 1          & 0.5        & 0.5        & 0.5        &            & 0.5        & 0.5         &             &             & 0           & 5                                                              & 55.5             & 13                  \\
(Gedeon et. al, 2019) \cite{paper_6}                            & ER          & 1          & 1          & 1          & 1          & 1          & 1          & 0.5        &            & 0          & 1           &             &             & 0.5         & 8                                                              & 80               & 6                  \\
(Wang et. al, 2019) \cite{paper_7}                            & SP          & 0.5        &            & 1          & 1          & 1          & 1          & 0.5        &            & 1          & 0.5         &             &             & 1           & 7.5                                                            & 83.3             & 18                 \\
(Busanelli et. al, 2019) \cite{paper_8}                            & SP          & 1          &            & 1          & 1          & 1          & 1          & 0.5        &            & 1          & 0.5         &             &             & 0.5         & 7.5                                                            & 83.3             & 8                  \\
(Rigazzi et. al, 2019) \cite{paper_9}                            & SP          & 1          &            & 1          & 1          & 0.5        & 1           & 0.5        &            & 0.5        & 0.5         &             &             & 1           & 7                                                              & 77.8             & 14                 \\

(Wang et. al, 2019) \cite{paper_10}                           & SP          & 1          &            & 0.5        & 1          & 0.5        & 1          & 0.5        &            & 0.5        & 0.5         &             &             & 1           & 6.5                                                            & 72.2             & 7                  \\

(Qu et. al, 2020) \cite{paper_11}                           & OP          & 1          &            &            &            &            &            &            & 1          & 0.5        & 1           & 0.5         & 1           & 1           & 6                                                              & 85.7             & 17                 \\
(Meixner et. al, 2019) \cite{paper_12}                           & SP          & 1          &            & 1          & 0.5        & 1          & 0.5        & 0.5        &            & 0.5        & 1           &             &             & 0.5         & 6.5                                                            & 72.2             & 10                  \\
(Buzachis and Villari,2018)\cite{paper_13}                           & VR          & 1          & 1          &            &            &            &            & 1          &            & 0.5        & 1           &             &             & 1           & 6.5                                                            & 92.86            & 11                 \\
(Samanta et. al,2019) \cite{paper_14}                           & ER          & 1          & 1          & 1          & 1          & 1          & 1          & 0.5        &            & 1          & 0.5         &             &             & 1           & 9                                                              & 90               & 24                 \\
(Xu et al.,2019) \cite{paper_15}                           & SP          & 1          &            & 1          & 0.5        & 1          & 1          & 0.5        &            & 0.5        & 1           &             &             & 1           & 7.5                                                            & 83.3             & 71                 \\
(Nguyen et al.,2019) \cite{paper_16}                           & SP          & 1          &            & 1          & 1          & 0.5        & 0.5        & 0.5        &            & 1          & 1           &             &             & 1           & 7.5                                                            & 83.3             & 5                  \\
(Pontes and Curry,2021) \cite{paper_17}                           & SP          & 0.5        &            & 0          & 1          & 0.5        & 0.5        & 0.5        &            & 0.5        & 0           &             &             & 0.5         & 4                                                              & 44.4             & 7                  \\
(Thramboulidis et al.,2019) \cite{paper_18}                           & VR          & 0.5        & 1          &            &            &            &            & 0.5        &            & 1          & 1           &             &             & 0.5         & 5                                                              & 71.4             & 36                 \\
(Tefera et al.,2019) \cite{paper_19}                           & OP          & 0.5        &            &            &            &            &            &            & 0.5        & 0          & 0.5         & 0.5         & 0.5         & 0.5         & 3                                                              & 43               & 8                  \\
(Harjula et al.,2019) \cite{paper_20}                           & SP          & 1          &            & 1          & 1          & 1          & 1          & 0.5        &            & 1          & 0.5         &             &             & 1           & 8                                                              & 89               & 30                 \\

(Nikouei et al., 2019) \cite{paper_21}                                & SP                              & 1          &            & 1          & 1          & 1          & 1          & 0.5        &            & 0.5        & 0.5         &             &             & 1           & 7.5                                                            & 83.3                 & 28                     \\
(Villari et al., 2017) \cite{paper_22}                                & SP                              & 1          &            & 0          & 0.5        & 0.5        & 1          & 0.5        &            & 0.5        & 1           &             &             & 1           & 6                                                              & 66.7                 & 28                     \\
(Islam et al., 2019) \cite{paper_23}                                & SP                              & 1          &            & 1          & 1          & 1          & 0.5        & 0.5        &            & 1          & 1           &             &             & 1           & 8                                                              & 88.9                 & 19                     \\
(Jimenez and Schelen, 2019) \cite{paper_24}                                & SP                              & 1          &            & 1          & 1          & 1          & 1          & 1          &            & 1          & 1           &             &             & 1           & 9                                                              & 100                  & 2                      \\
(Jin et al., 2021) \cite{paper_25}                                & SP                              & 1          &            & 1          & 1          & 0.5        & 0.5        & 0.5        &            & 1          & 1           &             &             & 1           & 7.5                                                            & 83.3                 & 9                      \\
(Leppanen et al., 2019) \cite{paper_26}                                & SP                              & 1          &            & 0.5        & 0.5        & 0.5        & 0.5        & 1          &            & 1          & 1           &             &             & 1           & 7                                                              & 77.8                 & 14                     \\
(Sattari et al., 2020) \cite{paper_27}                                & SP                              & 1          &            & 1          & 1          & 0.5        & 0.5        & 0.5        &            & 1          & 1           &             &             & 1           & 7.5                                                            & 83.3                 & 5                      \\
(Yang et al., 2021) \cite{paper_28}                                & ER                              & 0          & 1          & 1          & 1          & 1          & 1          & 0.5        &            & 1          & 0.5         &             &             & 1           & 8                                                              & 80                   & 9                      \\
(Valera, 2019) \cite{paper_29}                                & SP                              & 0.5        &            & 1          & 1          & 1          & 1          & 0.5        &            & 1          & 1           &             &             & 1           & 8                                                              & 88.9                 & 35                     \\
(Houmani et al., 2020) \cite{paper_30}                                & SP                              & 1          &            & 1          & 1          & 1          & 0.5        & 1          &            & 0.5        & 1           &             &             & 1           & 8                                                              & 88.9                 & 13                     \\
(Song and Tilevich, 2019) \cite{paper_31}                                & SP                              & 1          &            & 1          & 1          & 0.5        & 0.5        & 0.5        &            & 0.5        & 1           &             &             & 1           & 7                                                              & 77.8                 & 12 \\
(Zavodovski et al., 2019)\cite{paper_32}                                & SP                              & 0.5        &            & 1          & 1          & 0.5        & 0.5        & 1          &            & 0.5        & 1           &             &             & 1           & 7                                                              & 77.8                 & 12 \\
(Sánchez-Gallegos et al., 2020) \cite{paper_33}                                & ER                              & 0          & 1          & 1          & 1          & 1          & 1          & 0.5        &            & 1          & 0.5         &             &             & 1           & 8                                                              & 80                   & 20 \\
(Nakamura et al., 2020) \cite{paper_34}                                & SP                              & 1          &            & 1          & 1          & 0.5        & 0.5        & 0.5        &            & 1          & 1           &             &             & 1           & 7.5                                                            & 83.3                 & 7  \\
(Yan et al., 2021) \cite{paper_35}                                & ER                              & 0.5        & 1          & 1          & 1          & 1          & 1          & 1          &            & 1          & 1           &             &             & 1           & 9.5                                                            & 95                   & 18 \\
(Sriborrirux and Laortum, 2020) \cite{paper_36}                                & SP                              & 0.5        &            & 1          & 1          & 0.5        & 1          & 0.5        &            & 0.5        & 0.5         &             &             & 0.5         & 6                                                              & 66.7                 & 2  \\
(Laso et al., 2021) \cite{paper_37}                                & SP                              & 1          &            & 1          & 0.5        & 0.5        & 0.5        & 1          &            & 0.5        & 1           &             &             & 1           & 7                                                              & 77.8                 & 11 \\
(Jimenez and Schelen, 2020) \cite{paper_38}                                & SP                              & 1          &            & 1          & 1          & 1          & 1          & 0.5        &            & 0.5        & 0.5         &             &             & 1           & 7.5                                                            & 83.3                 & 6  \\

(Samanta et al., 2019) \cite{paper_39}                                & SP                              & 1          &            & 1          & 0.5        & 1          & 1          & 1          &            & 1          & 1           &             &             & 0.5         & 8                                                              & 88.9                 & 25                                     \\
(Santana et al., 2021) \cite{paper_40}                                & ER                              & 0.5        & 1          & 1          & 1          & 1          & 1          & 0.5        &            & 1          & 1           &             &             & 0.5         & 8.5                                                            & 85                   & 11                                     \\
(Ermolenko et al., 2021) \cite{paper_41}      &        SP                   & 0.5                               &           & 0.5          & 1          & 1          & 1          & 0.5          &           & 0.5          & 0.5          &            &            & 1           & 6.5           & 72.2                                                              & 4                   \\
(Javed et al., 2020) \cite{paper_42}                                & ER                              & 1          & 1          & 1          & 1          & 1          & 1          & 0.5        &            & 1          & 0.5         &             &             & 1           & 9                                                              & 90                   & 35                                     \\
(Bulej et al., 2021) \cite{paper_43}                                & SP                              & 0.5        &            & 1          & 1          & 1          & 1          & 0.5        &            & 1          & 1           &             &             & 1           & 8                                                              & 88.9                 & 10                                      \\
(Li and Gelbke, 2018) \cite{paper_44}                                & EX                              & 0.5        & 1          &            &            &            &            &            & 1          &            &             &             &             & 0.5         & 3                                                              & 75                   & 3                                      \\
(Li et al., 2018) \cite{paper_45}                                & VR                              & 1          & 1          &            &            &            & 1          & 0.5        &            & 0.5        & 1           &             &             & 1           & 6                                                              & 85.7                 & 13                                     \\
(Dai et al., 2021) \cite{paper_46}                           & ER          & 0.5        & 1          & 1          & 1          & 1          & 1          & 0.5        &            & 0.5        & 0.5                   &             &             & 0.5         & 7.5                                                            & 75               & 0                  \\
(Carrusca et al., 2020) \cite{paper_47}                           & VR          & 1          & 1          &            &            &            & 1          & 1          &            & 0.5        & \multicolumn{1}{c}{1} &             &             & 1           & 6.5                                                            & 92               & 5                  \\
(Mahajan et al., 2020) \cite{paper_48}                           & SP          & 1          &            & 1          & 0.5        & 1          & 1          & 0.5        &            & 0.5        & 0.5                   &             &             & 0.5         & 6.5                                                            & 72.2             & 1                  \\
(Deng et al., 2020) \cite{paper_49}                           & ER          & 1          & 1          & 1          & 0.5        & 1          & 1          & 1          &            & 1          & 1                     &             &             & 1           & 9.5                                                            & 95               & 5                  \\
\bottomrule
\end{tabular}
\end{table}


\begin{table}[h]
\small
\begin{tabular}{@{}llllllllllllllllll@{}}
\cmidrule(lr){3-15}
                                  &             & \multicolumn{13}{c}{\textbf{Quality Questions}}                                                                                                                            &                                                                &                  &                    \\ \midrule
\multicolumn{1}{c}{\textbf{Reference}} & \multicolumn{1}{l}{\textbf{RT}} & \textbf{1}           & \textbf{2}           & \textbf{3}           & \textbf{4}           & \textbf{5}           & \textbf{6}           & \textbf{7}           & \textbf{8}           & \textbf{9}           & \textbf{10}          & \textbf{11}          & \textbf{12}          & \textbf{13}          & \textbf{\begin{tabular}[c]{@{}c@{}}Total\\ Score\end{tabular}} & \textbf{Qual.}     & \textbf{Cit.} \\ \midrule

(Shah, 2018) \cite{paper_50}                           & SP          & 1          &            & 1          & 1          & 1          & 1          & 0.5        &            & 0.5        & 1                     &             &             & 1           & 8                                                              & 88.9             & 9                  \\
(Immich et al., 2018) \cite{paper_51}                           & SP          & 0.5        &            & 1          & 1          & 1          & 1          & 0.5        &            & 0.5        & 1                     &             &             & 1           & 7.5                                                            & 83.3             & 5                  \\
(Balteanu et al., 2020) \cite{paper_52}                           & SP          & 1          &            & 1          & 1          & 1          & 1          & 0.5        &            & 1          & 1                     &             &             & 0.5         & 8                                                              & 88.9             & 3                  \\
(Sanchez-Gallegos et al., 2020) \cite{paper_53}                           & ER          & 1          & 1          & 1          & 1          & 1          & 1          & 1          &            & 1          & 0.5                   &             &             & 1           & 9.5                                                            & 95               & 28                 \\
(Han et al., 2020) \cite{paper_54}                           & SP          & 0.5        &            & 1          & 0.5        & 1          & 1          & 0.5        &            & 0.5        & 0.5                   &             &             & 1           & 6.5                                                            & 72.2             & 4                  \\
(Pelle et al., 2021) \cite{paper_55}                           & SP          & 1          &            & 1          & 1          & 1          & 1          & 1          &            & 1          & 0.5                   &             &             & 1           & 8.5                                                            & 94.4             & 21                 \\
(Giannone et al., 2020) \cite{paper_56}                           & ER          & 0.5        & 1          & 1          & 1          & 1          & 1          & 0.5        &            & 1          & 0.5                   &             &             & 1           & 8.5                                                            & 85               & 10                  \\
(Lynn et al., 2020) \cite{paper_57}                           & OP          & 1          &            &            &            &            &            &            & 1          & 0.5        & 1                     & 1           & 1           & 1           & 6.5                                                            & 92.9             & 12                  \\
(Alam et al., 2018) \cite{paper_58}                           & SP          & 1          &            & 1          & 1          & 1          & 1          & 0.5        &            & 0          & 0.5                   &             &             & 1           & 7                                                              & 77.8             & 144                \\
(Carnevale et al., 2019) \cite{paper_59}                           & VR          & 1          & 1          &            &            &            & 0.5        & 1          &            & 1          & 1                     &             &             & 1           & 6.5                                                            & 92.9             & 21                 \\
(Falas et al., 2017) \cite{paper_60}                           & SP          & 1          &            & 1          & 0.5        & 1          & 0.5        & 1          &            & 0.5        & 0.5                   &             &             & 0.5         & 6.5                                                            & 72.2             & 5                  \\
(Ray et al., 2020) \cite{paper_61}                           & SP          & 0.5        &            & 1          & 1          & 1          & 1          & 1          &            & 1          & 1                     &             &             & 0           & 7.5                                                            & 83.3             & 18                 \\
(He et al., 2021) \cite{paper_62}                           & ER          & 0.5        & 1          & 1          & 1          & 1          & 1          & 1          &            & 1          & 1                     &             &             & 0.5         & 9                                                              & 90               & 18                 \\
(Paschke, 2016) \cite{paper_63}                           & SP          & 0.5        &            & 1          & 1          & 1          & 1          & 0.5        &            & 1          & 1                     &             &             & 0           & 7                                                              & 77.8             & 15                 \\
(D'Souza et al., 2019) \cite{paper_64}                           & SP          & 0.5        &            & 1          & 1          & 1          & 1          & 1          &            & 0.5        & 0.5                   &             &             & 1           & 7.5                                                            & 83.3             & 4                  \\
(He et al., 2019) \cite{paper_65}                           & SP          & 1          &            & 1          & 0.5        & 1          & 1          & 1          &            & 0.5        & 1                     &             &             & 0.5         & 7.5                                                            & 83.3             & 14                 \\
(Vaquero et al., 2019) \cite{paper_66}                           & OP          & 1          &            &            &            &            &            &            & 1          & 1          & 1                     & 1           & 1           & 0.5         & 6.5                                                            & 92.9             & 61                 \\

(Zhang et al., 2021) \cite{paper_67}                           & SP          & 1          &            & 1          & 0.5        & 1          & 1          & 0.5        &            & 0          & 0.5         &             &             & 0.5         & 6                                                              & 66.7             & 6                  \\
(Ezzeddine et al., 2018) \cite{paper_68}                           & SP          & 0.5        &            & 1          & 0.5        & 1          & 1          & 0.5        &            & 0.5        & 1           &             &             & 0.5                              & 6.5                                                                                 & 72.2                                  & 4                  \\
(Kathiravelu et al., 2019) \cite{paper_69}                           & SP          & 0.5        &            & 1          & 1          & 1          & 1          & 1          &            & 1          & 0.5         &             &             & 0.5                              & 7.5                                                                                 & 83.3                                  & 14                 \\
(Jin et al., 2020) \cite{paper_70}                           & ER          & 0.5        & 0.5        & 1          & 1          & 1          & 1          & 1          &            & 1          & 0.5         &             &             & 1                                & 8.5                                                                                 & 85                                    & 18                 \\
(Dai et al., 2019) \cite{paper_71}                           & SP          & 1          &            & 1          & 1          & 1          & 1          & 0.5        &            & 1          & 1           &             &             & 0.5                              & 8                                                                                   & 88.9                                  & 23                 \\
(Klaas et al., 2020) \cite{paper_72}                           & SP          & 0.5        &            & 1          & 1          & 1          & 1          & 0.5        &            & 1          & 0.5         &             &             & 0.5                              & 7                                                                                   & 77.8                                  & 2                  \\
(Liu et al., 2019) \cite{paper_73}                           & SP          & 0.5        &            & 1          & 1          & 0.5        & 1          & 0.5        &            & 0          & 0.5         &             &             & 0.5                              & 5.5                                                                                 & 61.1                                  & 14                 \\
(Tato et al., 2018) \cite{paper_74}                           & SP          & 1          &            & 1          & 1          & 1          & 1          & 0.5        &            & 0.5        & 1           &             &             & 0.5                              & 7.5                                                                                 & 83.3                                  & 18                 \\
(Khalyly et al., 2020) \cite{paper_75}                           & VR          & 1          &            &            &            &            & 1          & 0.5        &            & 1          & 0.5         &             &             & 0                                & 4.5                                                                                 & 64.3                                  & 0                  \\
(Tato et al., 2019) \cite{paper_76}                           & SP          & 1          &            & 1          & 1          & 1          & 1          & 0.5        &            & 0.5        & 1           &             &             & 0                                & 7                                                                                   & 77.8                                  & 11                  \\
(Gaggero et al., 2020) \cite{paper_77}                           & SP          & 0.5        &            & 1          & 1          & 1          & 1          & 0.5        &            & 0.5        & 0.5         &             &             & 0.5                              & 6.5                                                                                 & 72.2                                  & 1                  \\
(Calderon-Gomez et al., 2020) \cite{paper_78}                          & SP          & 0.5        &            & 1          & 1          & 1          & 1          & 1          &            & 1          & 0.5         &             &             & 0                                & 7                                                                                   & 77.8                                  & 20                 \\
(Qiao et al., 2018) \cite{paper_79}                           & VR          & 0.5        & 1          &            &            &            & 1          & 0.5        &            & 0          & 0.5         &             &             & 0.5                              & 4                                                                                   & 57.1                                  & 12                 \\
(Buzachis et al., 2019) \cite{paper_80}                           & SP          & 1          &            & 1          & 1          & 1          & 1          & 0.5        &            & 1          & 1           &             &             & 0.5                              & 8                                                                                   & 88.9                                  & 16                 \\
(Tourani et al., 2019) \cite{paper_81}                           & SP          & 0.5        &            & 1          & 1          & 1          & 1          & 0.5        &            & 0.5        & 0.5         &             &             & 0.5                              & 6.5                                                                                 & 72.2                                  & 9                  \\
(Souza et al., 2018) \cite{paper_82}                           & SP          & 1          &            & 1          & 1          & 1          & 1          & 1          &            & 1          & 1           &             &             & 0.5                              & 8.5                                                                                 & 94.4                                  & 9                  \\
(Song et al., 2020) \cite{paper_83}                           & SP          & 0.5        &            & 1          & 1          & 1          & 1          & 1          &            & 1          & 1           &             &             & 1                                & 8.5                                                                                 & 94.4                                  & 2                  \\
(Wang et al., 2021) \cite{paper_84}                           & SP          & 0.5        &            & 1          & 1          & 1          & 1          & 1          &            & 1          & 0.5         &             &             & 1                                & 8                                                                                   & 88.9                                  & 173                \\
(Arjomandy et al., 2021) \cite{paper_85}                           & VR          & 0.5        & 0.5        &            &            &            & 1          & 1          &            & 1          & 1           &             &             & 1                                & 5.5                                                                                 & 78.6                                  & 0                  \\
(Raffin et al., 2019) \cite{paper_86}                           & SP          & 0.5        &            & 1          & 1          & 1          & 0.5        & 0.5        &            & 0.5        & 0.5         &             &             & 1                                & 6.5                                                                                 & 72.2                                  & 0                  \\
(Mahmud and Toosi, 2021) \cite{paper_87}                           & SP          & 1          &            & 1          & 1          & 1          & 1          & 1          &            & 1          & 0.5         &             &             & 0.5                              & 8                                                                                   & 88.9                                  & 12                 \\

(Nadig et al., 2021) \cite{paper_88}                                & SP          & 0.5        &            & 1          & 1          & 1          & 1          & 0.5        &            & 0          & 0.5         &             &             & {0.5}         & {6}                                                              & {66.7}             & 4                  \\
(Seo et al., 2021) \cite{paper_89}                                & VR          & 0.5        & 1          &            &            &            & 1          & 0.5        &            & 0          & 0.5         &             &             & 1                                & 4.5                                                                                 & 64.3                                  & 0                  \\
(Simon et al., 2021) \cite{paper_90}                                & SP          & 0.5        &            & 1          & 0.5        & 1          & 1          & 1          &            & 0.5        & 1           &             &             & 0.5                              & 7                                                                                   & 77.8                                  & 0                  \\
(Pandey et al., 2021) \cite{paper_91}                                & SP          & 1          &            & 1          & 1          & 1          & 1          & 1          &            & 1          & 0.5         &             &             & 0.5                              & 8                                                                                   & 88.9                                  & 0                  \\
(Ouahabi et al., 2021) \cite{paper_92}                                & VR          & 0.5        & 1          &            &            &            & 1          & 1          &            & 1          & 0.5         &             &             & 1                                & 6                                                                                   & 85.7                                  & 7                  \\
(Ghosh et al., 2021) \cite{paper_93}                                & SP          & 0.5        &            & 1          & 1          & 1          & 1          & 0.5        &            & 0.5        & 0.5         &             &             & 1                                & 7                                                                                   & 77.8                                  & 7                  \\
(Islam et al., 2021) \cite{paper_94}                                & SP          & 1          &            & 1          & 0.5        & 1          & 1          & 0.5        &            & 1          & 1           &             &             & 1                                & 8                                                                                   & 88.9                                  & 15                 \\
(Fu et al., 2021) \cite{paper_95}                                & SP          & 1          &            & 1          & 1          & 1          & 1          & 1          &            & 1          & 1           &             &             & 1                                & 9                                                                                   & 100                                   & 20                 \\
(Lin et al., 2021) \cite{paper_96}                                & SP          & 0.5        &            & 1          & 1          & 1          & 1          & 0.5        &            & 1          & 1           &             &             & 0.5                              & 7.5                                                                                 & 83.3                                  & 3                  \\
(Ruan et al., 2021) \cite{paper_97}                                & SP          & 1          &            & 1          & 0.5        & 1          & 1          & 0.5        &            & 1          & 1           &             &             & 1                                & 8                                                                                   & 88.9                                  & 1                  \\
(Ullah et al., 2021) \cite{paper_98}                                & SP          & 1          &            & 1          & 1          & 1          & 1          & 1          &            & 1          & 1           &             &             & 1                                & 9                                                                                   & 100                                   & 15                 \\
(Xu et al., 2021) \cite{paper_99}                                & SP          & 1          &            & 1          & 1          & 0.5        & 1          & 1          &            & 1          & 1           &             &             & 1                                & 8.5                                                                                 & 94.4                                  & 6                  \\

\bottomrule
\end{tabular}
\end{table}

\begin{table}[h]
\small
\begin{tabular}{@{}llllllllllllllllll@{}}
\cmidrule(lr){3-15}
                                  &             & \multicolumn{13}{c}{\textbf{Quality Questions}}                                                                                                                            &                                                                &                  &                    \\ \midrule
\multicolumn{1}{c}{\textbf{Reference}} & \multicolumn{1}{l}{\textbf{RT}} & \textbf{1}           & \textbf{2}           & \textbf{3}           & \textbf{4}           & \textbf{5}           & \textbf{6}           & \textbf{7}           & \textbf{8}           & \textbf{9}           & \textbf{10}          & \textbf{11}          & \textbf{12}          & \textbf{13}          & \textbf{\begin{tabular}[c]{@{}c@{}}Total\\ Score\end{tabular}} & \textbf{Qual.}     & \textbf{Cit.} \\ \midrule

(Tian et al., 2021) \cite{paper_100}                               & VR                     & \multicolumn{1}{l}{1}   & \multicolumn{1}{l}{1} & \multicolumn{1}{l}{}  &                                &                                & 1                              & 0.5                            & \multicolumn{1}{l}{} & 1                              & 1                               & \multicolumn{1}{l}{} & \multicolumn{1}{l}{} & \multicolumn{1}{l}{0.5} & \multicolumn{1}{l}{6}                                          & \multicolumn{1}{l}{85.7} & 19                                     \\
(Kumara et al., 2021) \cite{paper_101}                               & SP                     & \multicolumn{1}{l}{0.5} & \multicolumn{1}{l}{}  & \multicolumn{1}{l}{1} & 1                              & 1                              & 1                              & 1                              & \multicolumn{1}{l}{} & 1                              & 1                               & \multicolumn{1}{l}{} & \multicolumn{1}{l}{} & \multicolumn{1}{l}{0.5} & \multicolumn{1}{l}{8}                                          & \multicolumn{1}{l}{88.9} & 9                                      \\
(Spillner et al., 2022)\cite{paper_102}                               & SP                     & \multicolumn{1}{l}{1}   & \multicolumn{1}{l}{}  & \multicolumn{1}{l}{1} & 1                              & 1                              & 1                              & 1                              & \multicolumn{1}{l}{} & 0.5                            & 1                               & \multicolumn{1}{l}{} & \multicolumn{1}{l}{} & \multicolumn{1}{l}{1}   & \multicolumn{1}{l}{8.5}                                        & \multicolumn{1}{l}{94.4} & 0                                      \\
(Chen and Liu, 2021) \cite{paper_103}                               & SP                     & 0.5                     &                       & \multicolumn{1}{l}{1} & 1                              & 1                              & 1                              & \multicolumn{1}{c}{0.5}        &                      & 1                              & \multicolumn{1}{c}{0.5}         &                      &                      & 1                       & 7.5                                                            & 83.3                     & 9                                      \\
(Miao et al., 2021) \cite{paper_104}                               & SP                     & 1                       &                       & 1                     & \multicolumn{1}{c}{0.5}        & 1                              & 1                              & 1                              &                      & 1                              & 1                               &                      &                      & 0.5                     & 8                                                              & 88.9                     & 0                                      \\
(Shen et al., 2021) \cite{paper_105}                               & SP                     &     0.5                    &                       &        1               & \multicolumn{1}{c}{1}           & \multicolumn{1}{c}{1}           & \multicolumn{1}{c}{1}           & \multicolumn{1}{c}{0.5}           &                      & \multicolumn{1}{c}{1}           & \multicolumn{1}{c}{0.5}            &                      &                      &         0.5                &       7                                                         &         77.8                 & 2                   \\
(Deng et al., 2021) \cite{paper_106}                               & SP                     & 1                       &                       & 1                     & 1                              & 1                              & 1                              & 0.5                            &                      & 1                              & 1                               &                      &                      & \multicolumn{1}{l}{1}   & 8.5                                                            & 94.4                     & 2                                      \\
(Chen and Liu, 2021) \cite{paper_107}                               & SP                     & \multicolumn{1}{l}{1}   &                       & \multicolumn{1}{l}{1} & 1                              & 1                              & 1                              & 1                              &                      & 1                              & 1                               &                      &                      & \multicolumn{1}{l}{1}   & 9                                                              & 100                      & 1                                      \\
(Iori et al., 2022) \cite{paper_108}                               & \multicolumn{1}{c}{VR} & 1                       & 1                     &                       & \multicolumn{1}{c}{}           & \multicolumn{1}{c}{}           & \multicolumn{1}{c}{1}          & \multicolumn{1}{c}{1}          &                      & \multicolumn{1}{c}{1}          & \multicolumn{1}{c}{1}           &                      &                      & 0.5                     & 6.5                                                            & 92.9                     & 0                                      \\
(Zilic et al., 2022) \cite{paper_109}                               & SP                     & 1                       &                       & \multicolumn{1}{l}{1} & 1                              & \multicolumn{1}{c}{0.5}        & 1                              & 1                              &                      & \multicolumn{1}{c}{0.5}        & 1                               &                      &                      & 1                       & 8                                                              & 88.9                     & 1                                      \\
(Das et al., 2022) \cite{paper_110}                               & SP                     &                  0.5       &                       &          1             & \multicolumn{1}{c}{1}           & \multicolumn{1}{c}{1}           & \multicolumn{1}{c}{1}           & \multicolumn{1}{c}{1}           &                      & \multicolumn{1}{c}{1}           & \multicolumn{1}{c}{0.5}            &                      &                      &            1             &                  8                                              &         88.9                 &                    1                    \\ \midrule
Average                           &             & \scriptsize $0.76$       & \scriptsize 0.96       & \scriptsize 0.95       & \scriptsize 0.90       & \scriptsize 0.88       & \scriptsize 0.90       & \scriptsize 0.70       & \scriptsize \scriptsize 0.84       & \scriptsize 0.74       & \scriptsize 0.77        & \scriptsize 0.86        & \scriptsize 0.95        & \scriptsize 0.75                             & \scriptsize 7.22                                                                                & \scriptsize 81.77                                 & \scriptsize 14.97              \\ \bottomrule

\end{tabular}
\end{table}

\bibliography{slr_references}
\end{document}